\documentclass[11pt,a4paper]{article}
\pdfoutput = 1
\usepackage{jheppub}
\usepackage{color}
\usepackage{amsmath,bm}
\usepackage{graphicx}

\newcommand{\as}{\alpha_{\mathrm{s}}}
\newcommand{\LA}{\mathrm{A}}
\newcommand{\LB}{\mathrm{B}}
\newcommand{\LF}{\mathrm{F}}
\newcommand{\LI}{\mathrm{I}}

\newcommand{\LR}{\mathrm{R}}

\newcommand{\LT}{\mathrm{T}}
\newcommand{\La}{\mathrm{a}}
\newcommand{\Lb}{\mathrm{b}}
\newcommand{\Lc}{\mathrm{c}}
\newcommand{\Lf}{\mathrm{f}}

\newcommand{\Lg}{\mathrm{g}}
\newcommand{\Lp}{\mathrm{p}}
\newcommand{\Ls}{\mathrm{s}}

\newcommand{\mpone}{{m+1}}

\newcommand{\GeV}{\ \mathrm{GeV}}

\definecolor{red}{rgb}{1,0,0}

\newcommand{\MSbar}{\overline {\text{MS}}}

\def\ket#1{\big|{#1}\big\rangle}
\def\bra#1{\big\langle{#1}\big|}
\def\brax#1{\big\langle{#1}}   
\def\<>#1{\big\langle{#1}\big\rangle}
\def\[]#1{\big[{#1}\big]}

\def\sket#1{\big|{#1}\big)}
\def\sbra#1{\big({#1}\big|}
\def\sbrax#1{\big({#1}}        

\newbox\charbox
\newbox\slabox
\def\s#1{{      
        \setbox\charbox=\hbox{$#1$}
        \setbox\slabox=\hbox{$/$}
        \dimen\charbox=\ht\slabox
        \advance\dimen\charbox by -\dp\slabox
        \advance\dimen\charbox by -\ht\charbox
        \advance\dimen\charbox by \dp\charbox
        \divide\dimen\charbox by 2
        \raise-\dimen\charbox\hbox to \wd\charbox{\hss/\hss}
        \llap{$#1$}
}}

\title{Parton distribution functions in the context of parton showers}

\author[a]{Zolt\'an Nagy}

\author[b]{and Davison E.\ Soper}

\affiliation[a]{
DESY\\
Notkestrasse 85\\
22607 Hamburg, Germany
}

\affiliation[b]{
Institute of Theoretical Science\\
University of Oregon\\
Eugene, OR  97403-5203, USA
}

\emailAdd{Zoltan.Nagy@desy.de}
\emailAdd{soper@uoregon.edu}

\abstract{
When the initial state evolution of a parton shower is organized according to the standard ``backward evolution'' prescription, ratios of parton distribution functions appear in the splitting probabilities. The shower thus organized evolves from a hard scale to a soft cutoff scale. At the end of the shower, one expects that only the parton distributions at the soft scale should affect the results. The other effects of the parton distributions should have cancelled. This means that the kernels for parton evolution should be related to the shower splitting functions. If the initial state partons can have non-zero masses, this requires that the evolution kernels cannot be the usual $\MSbar$ kernels. We work out what the parton evolution kernels should be to match the shower evolution contained in the parton shower event generator \textsc{Deductor}, in which the b and c quarks have non-zero masses.
}

\preprint{DESY 13-240}

\begin{document}
\maketitle

\section{Introduction}
 
In a companion paper \cite{deductor}, we have introduced a parton shower event generator, \textsc{Deductor} \cite{DeductorCode}, that is designed to be amenable to improved treatments of spin and color. This event generator is based on our earlier work \cite{NSI,NSII}. Methods for an improved treatment of spin are described in ref.~\cite{NSspin} and methods for an improved treatment of color\footnote{The current \textsc{Deductor} code implements the LC+ treatment for color described in ref.~\cite{NScolor}, but in ref.~\cite{deductor} we work only in the leading color (LC) approximation.} are described in ref.~\cite{NScolor}. This shower generator contains features that differ from other parton shower event generators even when one uses the leading color approximation and averages over spins, as we do in ref.~\cite{deductor}. Two of these features are important for this paper. 

The first feature is that we use a shower evolution variable defined by the virtuality in a splitting divided by the energy of the mother parton. In a second companion paper \cite{ShowerTime}, we argue that this choice is advantageous because, at leading order in QCD, it factors hard interactions from softer interactions at the amplitude level. 

The second feature is that we allow initial state partons to have nonzero mass. We regard up, down, and strange quarks to be effectively massless. The top quark is so heavy that we do not treat it as a possible constituent of the proton. This leaves the bottom and charm quarks, which do appear in the initial state as constituents of the proton. We take $m_\Lb$ and $m_\Lc$ to be non-zero. Do parton masses matter? We are interested in using parton shower evolution to examine what happens when there is a hard process with a momentum scale $Q$ of 100 GeV or more. At this scale, $m_\Lb$ and $m_\Lc$ do not matter. However, a b or c quark that participates in the hard interaction at scale $Q$ ultimately arose from an initial state $\Lg \to \Lb + \bar \Lb$ or $\Lg \to \Lc + \bar \Lc$ splitting. If the virtuality scale for this splitting was somewhere around the heavy quark mass, then the mass does matter.

These two features of the parton shower evolution have implications for the parton distribution functions used in the shower. This paper concerns these implications. 

The general analysis of ref.~\cite{NSI} gives the equations for parton shower evolution with a full quantum treatment of color and spin. The analysis in ref.~\cite{deductor} makes the leading color approximation and averages over spins. The issues of how masses enter the evolution equation for an initial state shower apply both with and without color and spin. Accordingly, in this paper we address these issues using the definitions of ref.~\cite{NSI} with full color and spin.

We begin the analysis of this paper in section \ref{sec:partonsandshowers} by outlining some of the important features of the shower evolution used in \textsc{Deductor} and then defining the general structure of the evolution equations needed for parton distributions used in the shower. This is not completely straightforward because of the presence of quark masses and because the shower evolution uses parton distribution functions at fixed shower time rather than fixed $\MSbar$ renormalization scale. In section \ref{sec:pertsplitting}, we review the definitions from refs.~\cite{NSI,NSII} of the operators that generate shower evolution and relate these to ``perturbative'' versions of these operators, which differ from the full versions by not containing factors of ratios of parton distribution functions. In section \ref{sec:findingP}, we use this analysis to argue that the kernels in the evolution equation for parton distributions must bear a simple relationship to the splitting kernels in the shower evolution operators. Some work is needed to derive the needed functions from the shower evolution operators of refs.~\cite{NSI,NSII}. This analysis is placed in an appendix \ref{sec:splittingfctns}. With the needed functions from shower evolution, one determines the part of the parton evolution kernels that involve splitting variables $z$ not equal to 1. There are $\delta(1-z)$ terms that we find in section \ref{sec:findingamma} by using flavor and momentum sum rules. We state the results for the parton evolution kernels including masses in section~\ref{sec:result}. If one starts parton evolution at a low scale $Q_{\rm fit}$ with fixed input distributions, then at a high scale the parton distributions defined with shower evolution will differ from those defined with $\MSbar$ evolution. In section \ref{sec:differencemassmakes}, we derive a lowest order perturbative relation for this difference. In section \ref{sec:pdfbehavior}, we display numerical results for the difference between shower parton evolution and $\MSbar$ parton evolution. In section \ref{sec:NLOaddition}, we record a modification at next-to-leading order to the parton evolution that is used in \textsc{Deductor}. We offer some concluding remarks in section \ref{sec:conclusions}.

\section{Parton evolution and shower evolution}
\label{sec:partonsandshowers}

Consider the following scenario. Two hadrons, A and B, collide to produce a final state system, for example a W boson plus a jet. The final state system has momentum $Q_0$. Now, the parton shower evolution simulates the development of the final state and also the development of the initial state. In each case, the development works from relatively hard interactions to softer interactions. In the case of the initial state, this means going backwards in physical time \cite{Sjostrand:1985xi, Gottschalk:1986bk}. 

At each stage in the shower, the incoming partons are defined to be on shell with zero momentum transverse to the beam directions. Of course, in the exact Feynman diagrams that describe the shower, the initial state partons are not exactly on shell. It is part of the shower approximation that we treat them as being on shell. It is also an approximation to treat the incoming partons as if they had zero transverse momenta. However, this approximation is not as drastic as it seems. At each initial state splitting, the momenta of the final state particles are adjusted as described in section 7.3 and appendix A of ref.~\cite{ShowerTime} to account for the recoil from the transverse momentum of the initial state splitting. For this reason, the transverse momentum of a Z-boson produced in the Drell-Yan process is correctly generated \cite{NSDrellYan}.

\subsection{Initial state parton splitting}
\label{sec:ISsplitting}

Let us look at the kinematics of initial state parton splitting. We define momenta $p_\LA$ and $p_\LB$ associated with the hadrons. These obey $(p_\LA + p_\LB)^2 = s$, but we do not take $p_\LA$ and $p_\LB$ to be exactly the hadron momenta. Rather, it is convenient to define $p_\LA$ and $p_\LB$ so that $p_\LA^2 = p_\LB^2 = 0$.\footnote{We never need the exact hadron momenta, but note here that in the case that both hadrons are protons, we have
$
p_\LA^{\rm exact} = \gamma p_\LA + [M_\Lp^2/(\gamma s)]\, p_B
$
and
$
p_\LB^{\rm exact} = [M_\Lp^2 /(\gamma s)]\, p_\LA +  \gamma p_B
$
where $\gamma = \left[1 + (1 - M_\Lp^2/s)^{1/2}\,\right]/2$. Thus $p_\LA$ and $p_\LB$ are close to the exact momenta of the incoming hadrons.} At any time in the shower, the incoming partons have momenta $p_\La$ and $p_\Lb$. These are defined to be on shell, with flavors $a$ and $b$, with masses $m(a)$ and $m(b)$, and with zero momentum transverse to the beam. We define momentum fractions $\eta_\La$ and $\eta_\Lb$ so that the momenta $p_\La$ and $p_\Lb$ are
\begin{equation}
\begin{split}
\label{eq:papbdef}
p_\La ={}& \eta_\La p_\LA 
+ \frac{m(a)^2}{\eta_\La \eta_\Lb s}\, \eta_\Lb p_\LB
\;,
\\
p_\Lb ={}& \eta_\Lb p_\LB 
+ \frac{m(b)^2}{\eta_\La \eta_\Lb s}\, \eta_\La p_\LA
\;.
\end{split}
\end{equation}

Now suppose that parton ``a'' splits in the sense of backwards evolution. Before the splitting, suppose that there were $m$ final state partons plus the two initial state partons. We denote momenta after the splitting by momentum vectors with hats, $\hat p$. The momentum of parton ``b'' remains the same: $\hat p_\Lb = p_\Lb$. Parton ``a'' after the splitting has a new momentum fraction $\hat \eta_a$ and possibly a new flavor $\hat a$:
\begin{equation}
\begin{split}
\label{eq:papbdef}
\hat p_\La ={}& \hat\eta_\La p_\LA 
+ \frac{m(\hat a)^2}{\hat\eta_\La \eta_\Lb s}\, \eta_\Lb p_\LB
\;,
\\
\hat p_\Lb ={}&  p_\Lb 
\;.
\end{split}
\end{equation}
The splitting creates a new final state parton with label $m+1$, flavor $\hat f_{m+1}$, and momentum $\hat p_{m+1}$. Parton $m+1$ is on shell: $\hat p_{m+1}^2 = m(f_{m+1})^2$ and typically has some transverse momentum. Momentum is not locally conserved in the splitting: $\hat p_\La - \hat p_{m+1} \ne p_\La$. Rather, we conserve momentum globally by making a small Lorentz transformation on the final state spectator partons: $\hat p_j = \Lambda p_j$ for $j = 1,\dots,m$. (See section 7.3 and appendix A of ref.~\cite{ShowerTime}.)

The momentum fraction splitting variable is $z = \eta_\La/\hat\eta_\La$. The spacelike virtuality in the splitting is 
\begin{equation}
\mu^2 = -[(\hat p_\La - \hat p_{m+1})^2 - m(a)^2]
\;.
\end{equation}

We divide the virtuality by $2 \eta_\La\,p_\LA\cdot Q_0$ to define the shower time $t$ of the splitting,\footnote{In ref.~\cite{ShowerTime}, we use the dimensionful variable $\Lambda^2 = Q_0^2 \exp(-t)$ to express the shower ordering definition, but in this paper it is more convenient to use the dimensionless variable $\exp(-t)$.}
\begin{equation}
\label{eq:showertime}
e^{-t} = \frac{\mu^2}{2 \eta_\La p_\LA\cdot Q_0}
=\frac{\mu^2}{\eta_\La\mu_\LA^2}
\;,
\end{equation}
where
\begin{equation}
\label{eq:muAdef}
\mu_\LA^2 = 2 p_\LA \cdot Q_0
\end{equation}
is a parameter that is fixed throughout the initial state shower. Thus $\mu_\LA^2$ is twice the energy in the hard scattering process times the energy of hadron A, both as measured in the c.m.\ frame of the hard scattering process.\footnote{Compare this to the scale variable $\zeta_\LA = {(2 p_\LA \cdot Q_0)^2}/{Q_0^2}$ used in ref.~\cite{JCCDES1981} to aid in factoring transverse momentum dependent parton distributions from the hard process.}

\subsection{The strong coupling}
\label{sec:alphas}

The probability for an initial state splitting is proportional to $\as$. What should be the argument of $\as$? We use $\as(\lambda_\LR k_\LT^2)$, where $\lambda_\LR \approx 0.4$ is given in eq.~(\ref{eq:lambdaR}) below and $k_\LT^2 = (1-z) \mu^2$. These choices are helpful for improving the summation of large logarithms arising from the emission of soft gluons \cite{NSDrellYan,lambdaR}. However, for the analysis of the evolution of parton distributions in this paper, it is more convenient to use $\as(\mu^2/z)$. These are related by
\begin{equation}
\as(\lambda_\LR (1-z) \mu^2) = 
\as(\mu^2/z) 
- \as(\mu^2/z)^2\, \beta_0 \log(\lambda_\LR z(1-z))
+ \cdots
\;.
\end{equation}
where $\beta_0 = {33 - 2 n_\Lf}/({12 \pi})$ is the first coefficient in the QCD $\beta$ function. If we used this order $\as^2$ correction in the analysis of this paper, it would suggest order $\alpha_s^2$ corrections to parton evolution. However, higher order corrections to the shower splitting function would also lead to order $\alpha_s^2$ corrections to parton evolution. We do not know what the shower splitting function should be beyond the leading order, so we ignore these corrections to parton evolution with one exception: since the $\beta_0 \log(\lambda_\LR)$ correction to parton evolution is so simple, we add it in section~\ref{sec:NLOaddition} below.

\subsection{The role of parton distributions}
\label{sec:pdfs}

What parton distribution function describes the mother parton at the time of the splitting?  We take it to be a function $f_{a/A}(\eta_\La,\mu^2)$. If all partons were massless, we could use the $\overline{\rm MS}$ definition \cite{CSpartons} of parton distribution functions. These functions, $f_{a/A}^{\overline{\rm MS}}(\eta_\La,\mu^2)$, obey the standard DGLAP evolution equations \cite{DGLAP}. However, the partons are not all massless, so $f_{a/A}$ is a possibly different function from $f_{a/A}^{\overline{\rm MS}}$. We assume that the first order evolution equation for $f_{a/A}$ has the form
\begin{equation}
\begin{split}
\label{eq:evolution1}
\frac{d}{d\log(\mu^2)}\,
f_{a/A}(\eta_{\La},\mu^{2})
={}& 
\sum_{\hat a} 
\int\!\frac{dz}{z}\
\frac{\as(\mu^2)}{2\pi} 
P_{a\hat a}(z,\mu^2/z)\
f_{\hat a/A}(\eta_{\La}/z,\mu^{2})
\;.
\end{split}
\end{equation}
When all of the partons are massless, $P_{a\hat a}(z,\mu^2/z)$ is the standard DGLAP kernel, which does not depend on the scale parameter in this case. When some quarks have masses and one uses $\overline{\rm MS}$ parton distribution functions, then one conventionally switches between an $(n-1)$ flavor scheme and an $n$ flavor scheme when $\mu^2$ becomes large enough. Specifically (working to order $\as$ in the evolution equations), if $m$ is the mass of one of the quarks $q$, then for $\mu^2 < m^2$ one sets $f_{q/A}^{\overline{\rm MS}}(\eta_{\La},\mu^{2}) = 0$, while for $\mu^2 > m^2$ one lets $f_{q/A}^{\overline{\rm MS}}(\eta_{\La},\mu^{2}) \ne 0$, with evolution determined by the normal DGLAP splitting functions with $f_{q/A}^{\overline{\rm MS}}(\eta_{\La},m^{2}) = 0$ as a boundary condition. Thus effectively the $g \to q$ splitting kernel is
\begin{equation}
\label{eq:MSbarkernel}
P_{qg}^{\overline{\rm MS}}(z,\mu^2/z)
=
T_\LR[1 - 2 z(1-z)]\,\Theta(\mu^2 > m^2)
\;.
\end{equation}
We will see in this paper that, with masses, we will need some extra terms that depend on the relevant squared parton mass $m^2$. The mass dependence will not be the same as in eq.~(\ref{eq:MSbarkernel}). It will be a convenient convention for us to take the second argument of $P_{a\hat a}$ to be $\mu^2/z$.

With our choice of shower time, the dimensionful variable $\mu_\LA^2 e^{-t}$ defines the shower time for an initial state splitting in hadron A. We have
\begin{equation}
\label{eq:showertimetomu}
\mu_\LA^2 e^{-t} = {\mu^2}/{\eta_\La}
\;.
\end{equation}
Inside of shower evolution, we use the function $f_{a/A}(\eta_{\La},\mu^{2})$ to describe the parton distribution, but with a different notation that emphasizes the separate roles of the momentum fraction $\eta_\La$ and the shower time $t$:
\begin{equation}
\label{eq:pdfrelationLO}
\tilde f_{a/A}\!\left(\eta_\La,\mu^2/\eta_\La\right)
=
f_{a/A}\!\left(\eta_\La,\mu^2\right)
\;.
\end{equation}
This function represents the probability to find a parton with flavor $a$ and momentum fraction $\eta_\La$ at shower time $t$ given by eq.~(\ref{eq:showertimetomu}).

Using eq.~(\ref{eq:pdfrelationLO}), the corresponding evolution equation for $\tilde f_{a/A}\!\left(\eta_\La,\mu^2/\eta_\La\right)$ is
\begin{equation}
\begin{split}
\label{eq:evolution2}
\frac{d\,\tilde f_{a/A}(\eta_{\La},\mu^{2}/\eta_\La)}{d\log(\mu^2)}
={}& 
\sum_{\hat a} 
\int\!\frac{dz}{z}\
\frac{\as(\mu^2)}{2\pi}
P_{a\hat a}\!\left(z,\mu^2/z\right)\,
f_{\hat a/A}(\eta_{\La}/z,\mu^2)
\\
={}&
\sum_{\hat a} 
\int\!\frac{dz}{z}\
\frac{\as(\mu^2)}{2\pi}
P_{a\hat a}\!\left(z,\mu^2/z\right)\,
\tilde f_{\hat a/A}(\eta_{\La}/z, z\mu^2/\eta_\La)
\;.
\end{split}
\end{equation}
This is the exact evolution equation for $\tilde f_{a/A}(\eta_{\La},\mu^{2}/\eta_\La)$. For our purpose of analyzing shower evolution, we will want to drop some contributions that are higher order in $\as$ so that we write
\begin{equation}
\begin{split}
\label{eq:evolution3}
\frac{d\,\tilde f_{a/A}(\eta_{\La},\mu^{2}/\eta_\La)}{d\log(\mu^2)}
={}&
\sum_{\hat a} 
\int\!\frac{dz}{z}\
\frac{\as(\mu^2/z)}{2\pi}
P_{a\hat a}\!\left(z,\mu^2/z\right)\,
\tilde f_{\hat a/A}(\eta_{\La}/z, \mu^2/\eta_\La)
+ {\cal O}(\as^2)
\;.
\end{split}
\end{equation}
On the right hand side of the equation, we have changed the scale argument of $\as$ from $\mu^2$ to $\mu^2/z$ and we have changed the scale argument of $\tilde f$ from $z\mu^2/\eta_{\La}$ to $\mu^2/\eta_{\La}$. Using the renormalization group equation for $\as$ and the evolution equation eq.~(\ref{eq:evolution2}) for $\tilde f$, we see that these scale changes correspond to higher order adjustments to the evolution equation, denoted by the  $+{\cal O}(\as^2)$ notation in eq.~(\ref{eq:evolution3}). Since we will be dealing with parton shower evolution only at leading order in $\as$, these higher order terms will not concern us.  We will find that when the partons have mass, terms beyond those of the customary DGLAP evolution kernel are needed in the evolution kernel $P_{a\hat a}(z,\mu^2/z)$ in eq.~(\ref{eq:evolution3}). These extra terms appear at {\em leading order} in $\as$.

The evolution kernel $P_{a\hat a}\!\left(z,\mu^2/z\right)$ is not an ordinary function but a distribution, with singular behavior as $z \to 1$. We can specify part of the structure of the kernel and write the same equation using ordinary functions by writing
\begin{equation}
\begin{split}
\label{eq:evolution4}
\frac{d\,\tilde f_{a/A}(\eta_{\La},\mu^{2}/\eta_{\La})}{d\log(\mu^2)}
={}&
\sum_{\hat a} 
\int_0^{1-}\!dz
\bigg\{
\frac{\as(\mu^2/z)}{2\pi}\,\frac{1}{z}
P_{a\hat a}\!\left(z,\mu^2/z\right)\,
\tilde f_{\hat a/A}(\eta_{\La}/z, \mu^2/\eta_{\La})
\\&-
\delta_{a\hat a}
\frac{\as(\mu^2)}{2\pi}
\left[
\frac{2 C_a}{1-z}-\gamma_a(\mu^2)
\right]
\tilde f_{a/A}(\eta_{\La}, \mu^2/\eta_{\La})
\bigg\}
\\&
+ {\cal O}(\as^2)
\;.
\end{split}
\end{equation}
Here the lower limit of the $z$-integration is $z=0$. However, we define $\tilde f_{a/A}(\eta_{\La}, \mu^2/\eta_{\La}) = 0$ for $\eta_\La > 1$, so that in the first term $\tilde f_{\hat a/A}(\eta_{\La}/z, \mu^2/\eta_{\La}) = 0$ unless $z > \eta_{\La}$. The upper limit is infinitesimally less than $z=1$. The kernel $P_{a\hat a}\!\left(z,\mu^2/z\right)$ in the first term is an ordinary function. We anticipate that for $\hat a = a$, $P$ has a singularity as $z \to 1$ of the form $2C_a/(1-z)$.  Here, as we will find later, $C_a$ is $C_\LF$ or $C_\LA$ for quarks and gluons, respectively. The same constant $C_a$ appears in the second term, so that the singular behavior is cancelled. There is also a term $\gamma_a$, which we allow to depend on quark masses and on $\mu^2$. We will have to determine $\gamma_a$.

\section{The perturbative splitting operators}
\label{sec:pertsplitting}

The shower evolution of ref.~\cite{NSI} is based on the evolution equation
\begin{equation}
\label{eq:rhoevolution}
\frac{d}{dt}\sket{\rho(t)} =
[{\cal H}_I(t) - {\cal V}(t)]
\sket{\rho(t)}
\;.
\end{equation}
Here $\sket{\rho(t)}$ represents the state of the system at shower time $t$ and ${\cal H}_I(t)$ and ${\cal V}(t)$ are operators on the space of states; ${\cal H}_I(t)$ describes splitting, increasing the number of partons by one, while ${\cal V}(t)$ describes virtual graphs and unresolved splittings, leaving the number of partons unchanged. See ref.~\cite{deductor} and ref.~\cite{NSI} for a more complete description.

The splitting operator ${\cal H}_I(t)$ contains a factor with a ratio of parton distribution functions. Specifically, suppose that we start with a basis state\footnote{Recall that we work with the quantum density operator in color and spin, so there are two quantum color states $\{c'\}_m$ and $\{c\}_m$ and two quantum spin states $\{s'\}_m$ and $\{s\}_m$.} $\sket{\{p,f,s',c',s,c\}_{m}}$ with $m$ final state partons. The partons have momenta $\{p\}_m = \{p_\La, p_\Lb, p_1, \dots, p_m\}$, flavors $\{f\}_m$, colors $\{c',c\}_m$ and spins  $\{s',s\}_m$. After applying ${\cal H}_I(t)$, we get a state with $m+1$ final state partons with new quantum numbers. The matrix element of ${\cal H}_I(t)$ has the form
\begin{equation}
\begin{split}
\label{eq:HtoHpert}
&\sbra{\{\hat p,\hat f,\hat s',\hat c',\hat s,\hat c\}_{m+1}}
   {\cal H}_I(t)
  \sket{\{p,f,s',c',s,c\}_{m}} 
\\&\qquad\qquad =
\frac{n_\Lc(a) n_\Lc(b)\,\eta_{\La}\eta_{\Lb}}
{n_\Lc(\hat a) n_\Lc(\hat b)\,\hat \eta_{\La}\hat \eta_{\Lb}}\
\frac{{\tilde f_{\hat a/A}(\hat \eta_{\La},\mu_\LA^2  e^{-t})
\tilde f_{\hat b/B}(\hat \eta_{\Lb},\mu_\LB^2  e^{-t})}}
{\tilde f_{a/A}(\eta_{\La},\mu_\LA^2  e^{-t})
\tilde f_{b/B}(\eta_{\Lb},\mu_\LB^2 e^{-t})}
\
\\&\qquad\qquad \quad\times
\sbra{\{\hat p,\hat f,\hat s',\hat c',\hat s,\hat c\}_{m+1}}
{\cal H}_I^{\rm pert}(t)
\sket{\{p,f,s',c',s,c\}_{m}}
\;.
\end{split}
\end{equation}
Here there is a ratio of parton distribution functions after the splitting to parton distribution functions before the splitting. That is because before the splitting, the probability for the system to be in the specified state is proportional to parton distribution functions $\tilde f_{a/A}(\eta_{\La},\mu_\LA^2  e^{-t})$ and $\tilde f_{b/B}(\eta_{\Lb},\mu_\LB^2 e^{-t})$. After the splitting the probability is proportional to parton distribution functions with the new variables. Thus we need to cancel the old parton distribution functions and introduce the new ones. Coming along with the parton distribution functions there is a ratio of kinematic factors $\eta_{\La}\eta_{\Lb}$ and there is a ratio of color factors $n_\Lc(a) n_\Lc(b)$, where $n_\Lc(a)$ is the number of colors of a parton with flavor $a$. The rest of the matrix element of ${\cal H}_I(t)$, denoted here as a matrix element of a new operator ${\cal H}_I^{\rm pert}(t)$, is rather complicated but contains no factors of parton distribution functions.

It is precisely the ratio of parton distribution functions in eq.~(\ref{eq:HtoHpert}) that interests us in this paper. This ratio is standard in modern parton shower event generators and it is needed for an efficient generation of parton splittings. However, there is a sense in which a dependence on parton distribution functions should not be there. The very splittings described in ${\cal H}_I^{\rm pert}(t)$ are the splittings that generate the evolution of the parton distribution functions. Thus we should not need parton distribution functions to describe the splittings. The only parton distribution functions that we should need consist of one factor of parton distributions at the low virtuality end of the parton shower. Indeed, roughly this idea was present from the beginning of the development of parton showers with backwards evolution \cite{Sjostrand:1985xi,Gottschalk:1986bk}. The formulation used in \textsc{Deductor} follows most closely that of ref.~\cite{Sjostrand:1985xi}.

In order to investigate this idea, let us define an operator ${\cal F}(t)$ that multiplies by the parton distribution factor that relates the cross section to a squared matrix element,
\begin{equation}
\begin{split}
\label{eq:Fdef}
{\cal F}(t)&\sket{\{p,f,s',c',s,c\}_{m}}\\ &= 
\frac{\tilde f_{a/A}(\eta_{\La},\mu_\LA^2  e^{-t})
\tilde f_{b/B}(\eta_{\Lb}, \mu_\LB^2 e^{-t})}
{4n_\Lc(a) n_\Lc(b)\,4\eta_{\La}\eta_{\Lb}p_\LA\!\cdot\!p_\LB}\
\sket{\{p,f,s',c',s,c\}_{m}} 
\;.
\end{split}
\end{equation}
Then the operator ${\cal H}^{\rm pert}_I(t)$ defined in eq.~(\ref{eq:HtoHpert}) is
\begin{equation}
\begin{split}
\label{eq:HpertfromF}
{\cal H}^{\rm pert}_I(t)
 ={}& 
{\cal F}(t)^{-1}
{\cal H}_I(t)
{\cal F}(t)
\;.
\end{split}
\end{equation}

How should we define the corresponding operator ${\cal V}^{\rm pert}(t)$? To find out, first define a shower state vector $\sket{\rho_{\rm pert}(t)}$ that has the parton distribution factor removed:
\begin{equation}
\label{eq:rhopertdef}
\sket{\rho(t)} = {\cal F}(t) \sket{\rho_{\rm pert}(t)}
\;.
\end{equation}
The evolution equation for $\sket{\rho_{\rm pert}(t)}$ can be determined from the evolution equation (\ref{eq:rhoevolution}) for $\sket{\rho(t)}$.  We have
\begin{equation}
\begin{split}
\left[\frac{d}{dt}\,{\cal F}(t)\right]
\sket{\rho_{\rm pert}(t)}
+ {\cal F}(t) \frac{d}{dt} \sket{\rho_{\rm pert}(t)}
 ={}& 
[{\cal H}_I(t) - {\cal V}(t)]
{\cal F}(t)
\sket{\rho_{\rm pert}(t)}
\;,
\end{split}
\end{equation}
so
\begin{equation}
\begin{split}
 \frac{d}{dt} \sket{\rho_{\rm pert}(t)}
 ={}& 
 {\cal F}(t)^{-1}
[{\cal H}_I(t) - {\cal V}(t)]
{\cal F}(t)
\sket{\rho_{\rm pert}(t)}
\\ & 
- {\cal F}(t)^{-1}\left[\frac{d}{dt}\,{\cal F}(t)\right]
\sket{\rho_{\rm pert}(t)}
\;.
\end{split}
\end{equation}
We can write this as
\begin{equation}
\label{eq:rhopertevolution}
\frac{d}{dt}\sket{\rho_{\rm pert}(t)} =
[{\cal H}^{\rm pert}_I(t) - {\cal V}^{\rm pert}(t)]
\sket{\rho_{\rm pert}(t)}
\;.
\end{equation}
Here ${\cal H}^{\rm pert}_I(t)$ is given in eq.~(\ref{eq:HpertfromF}) and ${\cal V}^{\rm pert}(t)$ is
\begin{equation}
\begin{split}
\label{eq:VpertfromF}
{\cal V}^{\rm pert}(t)
 ={}& 
{\cal V}(t)
+ {\cal F}(t)^{-1}\left[\frac{d}{dt}\,{\cal F}(t)\right]
\;.
\end{split}
\end{equation}
Here we have noted that ${\cal F}(t)$ commutes with ${\cal V}(t)$ since ${\cal V}(t)$ does not change momenta or flavors.

Now we can make a couple of observations. First, the starting value of $\sket{\rho_{\rm pert}(t)}$ at the time $t_0$ that corresponds to the hard interaction does not contain parton distribution functions because we removed this factor from $\sket{\rho_{\rm pert}(t)}$. Second, if we let $\sket{\rho_{\rm pert}(t)}$ evolve to some late shower time $t_\Lf$, then we can recover the full shower state at $t_\Lf$ using
\begin{equation}
\label{eq:rhopertattf}
\sket{\rho(t_\Lf)} = {\cal F}(t_\Lf) \sket{\rho_{\rm pert}(t_\Lf)}
\;.
\end{equation}
Thus $\sket{\rho(t_\Lf)}$ contains the proper product of parton distribution functions as long as $\sket{\rho_{\rm pert}(t_\Lf)}$, like $\sket{\rho_{\rm pert}(t_0)}$, does not depend on parton distribution functions. This means that the evolution from $t_0$ to $t_\Lf$ should not have introduced any dependence on parton distribution functions. Now, the operator ${\cal H}^{\rm pert}_I(t)$ in the evolution equation~(\ref{eq:rhopertevolution}) for $\sket{\rho_{\rm pert}(t)}$ does not contain any parton distribution functions by construction. However, in eq.~(\ref{eq:VpertfromF}) for ${\cal V}^{\rm pert}(t)$, the operator ${\cal V}(t)$ does contain explicit parton distribution function factors. Additionally, ${\cal F}(t)^{-1}d{\cal F}(t)/dt$ contains parton distribution functions. Because of the differentiation with respect to $t$, it also contains the evolution kernel for the parton distribution functions. Thus, what needs to happen is that the evolution kernel for the parton distribution functions has the right form compared to the functions in ${\cal V}(t)$ so that the dependence on parton distribution functions cancels between the two terms in eq.~(\ref{eq:VpertfromF}), at least after applying suitable kinematic approximations that correspond to the parton splittings in the shower being approximately collinear or soft. This is the issue that we will investigate in the following sections.

\section{Shower kinematics}

We will want to examine the evolution of ${\cal V}^{\rm pert}(t)$. For this purpose, we need some kinematic variables for the initial state shower. 

At shower time $t$, an initial state parton from hadron A with momentum fraction $\eta_\La$ can become a new initial state parton with momentum fraction $\hat\eta_\La$ with the emission of a new final state parton with momentum $\hat p_{m+1}$. The ratio of momentum fractions is $\eta_\La/\hat\eta_\La = z$.

It is useful to define a dimensionless virtuality variable
\begin{equation}
y
= -\frac{(\hat p_\La - \hat p_{m+1})^2 - m(a)^2}{\eta_\La \eta_\Lb s}
= \frac{\mu^2}{\eta_\La \eta_\Lb s}
= \frac{\mu_\LA^2}{\eta_\Lb s}\,e^{-t}
\;.
\end{equation}
That is, $y$ is the virtuality in the splitting divided by the total current squared c.m.\ energy of the colliding partons, $\eta_\La \eta_\Lb s$. At the first initial state splitting, $y$ is much smaller than 1 as long as the first splitting is close to being collinear or soft. At each subsequent initial state splitting, $t$ is larger than in the previous splitting and $\eta_b$ is the same or larger. Thus $y$ gets smaller at each splitting. For this reason, in a parton shower it is a good approximation to assume $y \ll 1$. 

It is also useful to define a dimensionless mass squared variable
\begin{equation}
\label{eq:nufdef}
\nu(f) = 
\frac{m(f)^2}{\eta_\La \eta_\Lb s}
\;,
\qquad f = a,\hat a, \text{ or } \hat f_{m+1}
\;.
\end{equation}
For u, d, and s quarks we can take $\nu(f) = 0$. For c and b quarks, $\nu(f) \ne 0$. However, we are interested in hard processes for which the scale is much greater than squared quark masses:\footnote{If we wanted to consider b-quark production at the LHC with the b-quark transverse momentum similar to the b-quark mass, then we would not have $\nu_\Lb \ll 1$. But then, we should not let the b quark be an active parton that is treated as a constituent of the proton.} $Q_0^2 \gg m(f)^2$. Thus $\nu(f) \ll 1$ at the start of the shower. At each subsequent initial state splitting, $\eta_a$ and $\eta_b$ are the same or larger than they were at the start of the shower. For this reason, in a parton shower it is a good approximation to assume $\nu(f) \ll 1$.

\section{Determining $P_{a\hat a}\!\left(z,\mu^2/z\right)$ at finite $z$}
\label{sec:findingP}

As argued in the section \ref{sec:pertsplitting}, we want to arrange that the virtual splitting operator 
\begin{equation*}
\begin{split}
{\cal V}^{\rm pert}(t)
 ={}& 
{\cal V}(t)
+ {\cal F}(t)^{-1}\left[\frac{d}{dt}\,{\cal F}(t)\right]
\end{split}
\end{equation*}
does not involve parton distribution functions after suitable kinematic approximations are applied.

Let us look at the second term in ${\cal V}^{\rm pert}(t)$. Our partonic basis states are eigenfunctions of this operator:
\begin{equation}
\begin{split}
\label{eq:Vperteigenvalues}
{\cal F}(t)^{-1}\left[\frac{d}{dt}\,{\cal F}(t)\right]&
\sket{\{p,f,s',c',s,c\}_{m}}
\\ ={}& 
[\lambda^{\cal F}_{\La}(a,\eta_\La,t)
+ \lambda^{\cal F}_{\Lb}(b,\eta_\Lb,t)]
\sket{\{p,f,s',c',s,c\}_{m}} 
\;,
\end{split}
\end{equation}
where
\begin{equation}
\lambda^{\cal F}_{\La}(a,\eta_\La,t)
=
\frac{\frac{d}{dt}\,\tilde f_{a/A}(\eta_\La,\mu_\LA^2  e^{-t})}
{\tilde f_{a/A}(\eta_\La,\mu_\LA^2  e^{-t})}
\end{equation}
with a corresponding expression for $\lambda^{\cal F}_{\Lb}$. Using the parton evolution equation (\ref{eq:evolution4}), this is
\begin{equation}
\begin{split}
\label{eq:Feigenvalue}
\lambda^{\cal F}_{\La}(a,\eta_\La,t)
={}&
-\sum_{\hat a} 
\int_0^{1-}\!dz\
\bigg\{
\frac{\as(\mu^2/z)}{2\pi}\,\frac{1}{z}
P_{a\hat a}\!\left(z,\frac{\mu^2}{z}\right)
\,
\frac{\tilde f_{\hat a/A}(\eta_{\La}/z, \mu^2/\eta_\La)}
{\tilde f_{a/A}(\eta_{\La}, \mu^2/\eta_\La)}
\\&-
\delta_{a\hat a}
\frac{\as(\mu^2)}{2\pi}
\left[
\frac{2 C_a}{1-z}-\gamma_a\!\left(\mu^2\right)
\right]
\bigg\}
+ {\cal O}(\as^2)
\;.
\end{split}
\end{equation}
The first term in $\lambda^{\cal F}_{\La}$ involves a ratio of parton distribution functions. We need to somehow make this term go away.

We now look at the first term in ${\cal V}^{\rm pert}(t)$, namely ${\cal V}(t)$. 
This operator has a contribution for each initial state or final state parton,
\begin{equation}
{\cal V}(t)\sket{\{p,f,s',c',s,c\}_{m}} 
= \left[
{\cal V}_\La(t)
+ {\cal V}_\Lb(t)
+ \sum_{l=1}^m {\cal V}_l(t)
\right]\sket{\{p,f,s',c',s,c\}_{m}}
\;.
\end{equation}
The contributions ${\cal V}_l(t)$ from final state partons do not contain any factors of ratios of parton distributions, so we can ignore these terms. There are two choices for initial state partons, but ${\cal V}_\Lb(t)$ has the same structure as ${\cal V}_\La(t)$, so we can concentrate on ${\cal V}_\La(t)$. 

The operator ${\cal V}_\La(t)$ contains two kinds of terms,
\begin{equation}
{\cal V}_\La(t)\sket{\{p,f,s',c',s,c\}_{m}} = 
\left[
{\cal V}_{\La\La}(t)
+ \sum_{k \ne \La}{\cal V}_{\La k}(t)
\right]\sket{\{p,f,s',c',s,c\}_{m}}
\;.
\end{equation}
The term ${\cal V}_{\La\La}(t)$ is derived from parton splittings in which parton ``a'' splits in the ket state and in the conjugate bra state. We will return to it shortly. The terms ${\cal V}_{\La k}(t)$ are derived from interference graphs in which parton ``a'' emits a gluon in the ket state but a different parton, $k$, emits the gluon in the bra state or in which parton ``a'' emits a gluon in the bra state and parton $k$ emits the gluon in the ket state. The action of ${\cal V}_{\La k}(t)$ on a basis state has a simple form,
\begin{equation}
{\cal V}_{\La k}(t)\sket{\{p,f,s',c',s,c\}_{m}}
= 
\sum_{\bar c'\bar c}\lambda^{\cal V}_{\La k}(\{p,f\}_{m},t)^{c'c}_{\bar c'\bar c}
\sket{\{p,f,s',\bar c',s,\bar c\}_{m}}
\;.
\end{equation}
That is, ${\cal V}_{\La k}(t)$ leaves momenta, flavors, and spins unchanged but acts according to a matrix in color space. The color-space matrix has the form
\begin{equation}
\begin{split}
\label{eq:VeigenvalueSoft}
\lambda^{\cal V}_{\La k}(\{p,f\}_{m},t)^{c'c}_{\bar c'\bar c}
={}& 
\int_0^{1-}\!dz\
\frac{\as(\mu^2/z)}{2\pi}
\frac{1}{z}\,
g_{\La k}\!\left(z,{\mu^2}/{z},\{p,f\}_{m}\right)^{c'c}_{\bar c'\bar c}\
\frac{\tilde f_{a/A}(\eta_{\La}/z, \mu^2/\eta_\La)}
{\tilde f_{a/A}(\eta_\La,\mu^2/\eta_\La)}
\;.
\end{split}
\end{equation}
(We choose the scale arguments of $\as$ and $\tilde f_{a/A}$ as discussed in sections~\ref{sec:alphas} and \ref{sec:pdfs}.)
 
We need to understand the structure of the function $g_{a k}$. It contains a color matrix that need not concern us here and a function $A_{\La k}$ that defines how much of the interference graph is attributed to a splitting of parton ``a'' and how much is attributed to a splitting of parton $k$. The only important feature of $A_{\La k}$ is that it is everywhere finite. The essential factor in $g_{a k}$ is the eikonal approximation to the Feynman graph,
\begin{equation}
\frac{\hat p_\La \cdot D(\hat p_{m+1}) \cdot \hat p_k}{\hat p_{m+1}\cdot \hat p_\La\, \hat p_{m+1}\cdot \hat p_k}
\;,
\end{equation}
where $D(\hat p_{m+1})^{\mu\nu}$ is the polarization sum for the emitted gluon in Coulomb gauge. This factor is singular in the region of wide angle soft gluon emission, but it is not singular when gluon $m+1$ becomes collinear with $p_\La$ or $p_k$. Now, at small $y$, we integrate over $z$. There are three integration regions to consider: the collinear region $y \ll (1-z) \sim 1$; the soft region $y \sim (1-z) \ll 1$; and the intermediate region, $y \ll (1-z) \ll 1$. However, only the soft region $y \sim (1-z) \ll 1$ is important.  For that reason, in eq.~(\ref{eq:VeigenvalueSoft}) we can approximate $z$ by 1 in the parton distribution functions (and also elsewhere). This gives
\begin{equation}
\begin{split}
\label{eq:VeigenvalueSoftmod}
\lambda^{\cal V}_{\La k}(\{p,f\}_{m},t)^{c'c}_{\bar c'\bar c}
\sim{}&
\sum_{\hat a} 
\int_0^{1-}\!dz\
\frac{\as(\mu^2)}{2\pi}\
g_{\La k}\!\left(z, {\mu^2}/{z} ,\{p,f\}_{m}\right)^{c'c}_{\bar c'\bar c}
\;.
\end{split}
\end{equation}
There is no factor of the ratio of parton distribution functions in the contribution to ${\cal V}(t)$ from $\lambda^{\cal V}_{\La k}(t)$, so this contribution does not help to cancel the ratio of parton distribution functions in eq.~(\ref{eq:Feigenvalue}).

To avoid confusion, let us note that the functions $g_{\La k}$ are important in the parton shower. They help determine the part of the development of the parton shower that comes from soft gluon emissions. However, they do not play a role in the present analysis because, in the limit of small $y$, they do not multiply parton distribution functions. 

Next, we examine ${\cal V}_{\La\La}(t)$, which contains the functions that we will really need. The states $\sket{\{p,f,s',c',s,c\}_{m}}$ are eigenvectors of ${\cal V}_{\La\La}(t)$:
\begin{equation}
\label{eq:Veigenvaluedef}
{\cal V}_{\La\La}(t)\sket{\{p,f,s',c',s,c\}_{m}}
= \lambda^{\cal V}_{\La\La}(\{p,f\}_{m},t)\sket{\{p,f,s',c',s,c\}_{m}}
\;.
\end{equation}
The eigenvalue $\lambda^{\cal V}_{\La\La}$ is made of a factor of $\as$, a ratio of parton distribution functions, and a certain function $g_{a \hat a}$:
\begin{equation}
\begin{split}
\label{eq:Veigenvalue}
\lambda^{\cal V}_{\La\La}(\{p,f\}_{m},t)
={}&
\sum_{\hat a} 
\int_0^{1-}\!dz\
\frac{\as(\mu^2/z)}{2\pi}\,
\frac{1}{z}\,
g_{a\hat a}\!\left(z,{\mu^2}/{z},\{p,f\}_{m}\right)\,
\frac{\tilde f_{\hat a/A}(\eta_{\La}/z, \mu^2/\eta_\La)}
{\tilde f_{a/A}(\eta_\La,\mu^2/\eta_\La)}
\;.
\end{split}
\end{equation}
Again, we choose the scale arguments of $\as$ and $\tilde f_{a/A}$ as discussed in sections~\ref{sec:alphas} and \ref{sec:pdfs}. The function $g$ corresponds to parton splittings in which $a$ is the flavor index of the parton after the splitting and $\hat a$ is the flavor index of the parton before the splitting (thinking of the process going forward in time). We need to understand the structure of the functions $g_{a\hat a}$. 

The function $g_{a\hat a}$ is rather complicated, but it is simple when $y \ll 1$ and $\nu(f) \ll 1$  for $f = a,\hat a,b$, with no requirement on the ratio of $y$ to $\nu(f)$. This function appears inside an integration over $z$ and both the regions of finite $(1-z)$ (the collinear region) and of $(1-z) \sim y \ll 1$ (the wide angle soft region) are important in the integration. The intermediate region, $(1-z) \ll y \ll 1$, is also important. In the wide angle soft region, the structure of $g_{a\hat a}$ is particularly simple,
\begin{equation}
\label{eq:gsoftregion}
g_{a\hat a} \sim 
\delta_{a\hat a}2 C_a
\left[\frac{1}{1-z}- \frac{y}{(1-z)^2}\right]\Theta((1-z) > y)
\;,
\end{equation}
where
\begin{equation}
\label{eq:Cadef}
C_a = 
\begin{cases}
C_\LF  & a\ne \Lg \\
C_\LA  & a = \Lg 
\end{cases}
\;.
\end{equation}
The constraint $(1-z) > y$ arises from the kinematics. It is useful to write this as
\begin{equation}
g_{a\hat a} \sim 
\delta_{a\hat a}\,\frac{2C_a}{1-z}
-
\delta_{a\hat a}\,\frac{2C_a}{1-z}
\,\Theta((1-z) < y)
-
\delta_{a\hat a}\,\frac{2C_a\, y}{(1-z)^2}
\,\Theta((1-z) > y)
\;.
\end{equation}
With this notation, we find that when $y \ll 1$ and $\nu(f) \ll 1$ we have
\begin{equation}
\begin{split}
\label{eq:gtoG}
g_{a\hat a}\!\left(z,{\mu^2}/{z},\{p,f\}_{m}\right)
\sim{}& 
G_{a\hat a}\!\left(z,{\mu^2}/{z}\right)
\\&
- \delta_{a\hat a}\,\frac{2C_a}{1-z}
\left[\Theta((1-z)<y)
+
\frac{y\, \Theta((1-z) > y)}{(1-z)}\right]
\;.
\end{split}
\end{equation}
The function $G_{a\hat a}$ is an ordinary function of its arguments and is independent of the arguments in $\{p,f\}_{m}$ other than $\eta_\La$ and $a$. We leave a detailed determination of this function for the appendix \ref{sec:splittingfctns}. 

Inserting eq.~(\ref{eq:gtoG}) into eq.~(\ref{eq:Veigenvalue}), we have
\begin{equation}
\begin{split}
\label{eq:Veigenvalue2}
\lambda^{\cal V}_{\La\La}(\{p,f\}_{m},t)
={}&
\sum_{\hat a} 
\Bigg\{
\int_0^{1-}\!dz\
\frac{\as(\mu^2/z)}{2\pi}
\bigg[
\frac{1}{z}\,
G_{a\hat a}\!\left(z,{\mu^2}/{z}\right)\,
\frac{\tilde f_{\hat a/A}(\eta_{\La}/z, {\mu^2}/{\eta_\La})}
{\tilde f_{a/A}(\eta_\La,{\mu^2}/{\eta_\La})}
\\&\qquad
- \delta_{a\hat a}\,\frac{1}{z}\,\frac{2C_a}{1-z}
\left[\Theta((1-z)<y)
+
\frac{y\, \Theta((1-z) > y)}{(1-z)}\right]
\\&\hskip 2 cm\times
\frac{\tilde f_{\hat a/A}(\eta_{\La}/z, {\mu^2}/{\eta_\La})}
{\tilde f_{a/A}(\eta_\La,{\mu^2}/{\eta_\La})}
\bigg]\
\\&
+ {\cal O}(y,\nu(a),\nu(\hat a))
\Bigg\}
\;.
\end{split}
\end{equation}
Now, consider the second term on the right hand side of eq.~(\ref{eq:Veigenvalue2}). The only region of the $z$ integration that matters for $y \ll 1$ is the region $(1-z) \lesssim y$. We can make further use of the approximation $y \ll 1$ by replacing $z$ by 1 in the factor $1/z$, the argument of $\as$, and, more importantly, in the argument of the parton distribution functions. This removes the ratio of parton distribution functions from this term. We are left with
\begin{equation}
\begin{split}
\label{eq:Veigenvalue3}
\lambda^{\cal V}_{\La\La}(\{p,f\}_{m},t)
={}&
\sum_{\hat a} 
\int_0^{1-}\!dz\
\Bigg\{
\frac{\as(\mu^2/z)}{2\pi}
\frac{1}{z}\,
G_{a\hat a}\!\left(z,{\mu^2}/{z}\right)\,
\frac{\tilde f_{\hat a/A}(\eta_{\La}/z, {\mu^2}/{\eta_\La})}
{\tilde f_{a/A}(\eta_\La,{\mu^2}/{\eta_\La})}
\\&\qquad
- \delta_{a\hat a}\,
\frac{\as(\mu^2)}{2\pi}
\frac{2C_a}{1-z}
\left[\Theta((1-z)<y)
+
\frac{y\, \Theta((1-z) > y)}{(1-z)}\right]
\\&
+ {\cal O}(y,\nu(a),\nu(\hat a))
+ {\cal O}(\as^2)
\Bigg\}
\;.
\end{split}
\end{equation}
Combining eqs.~(\ref{eq:Feigenvalue}) and (\ref{eq:Veigenvalue3}) we have
\begin{equation}
\begin{split}
\label{eq:VFeigenvalue}
\lambda^{\cal F}_{\La}(a,\eta_\La,t)&
+ \lambda^{\cal V}_{\La\La}(\{p,f\}_{m},t)
\\
={}&
\sum_{\hat a} 
\Bigg\{
\int_0^{1-}\!dz \bigg[
\frac{\as(\mu^2/z)}{2\pi}\,\frac{1}{z}\,
\frac{\tilde f_{\hat a/A}(\eta_{\La}/z, \mu^2/\eta_\La)}
{\tilde f_{a/A}(\eta_\La, \mu^2/\eta_\La)}
\\&\qquad \times
\left\{
G_{a\hat a}\!\left(z,\mu^2/z\right)
-
P_{a\hat a}\!\left(z,\mu^2/z\right)
\right\}
\\&\qquad
+ \delta_{a\hat a}\,
\frac{\as(\mu^2)}{2\pi}\,
\frac{2C_a}{1-z}\left[\Theta((1-z) > y)
-
\frac{y\, \Theta((1-z) > y)}{(1-z)}\right]
\\&
-
\delta_{a\hat a}
\frac{\as(\mu^2)}{2\pi}\,
\gamma_a\!\left(\mu^2\right)
\bigg]
\\&
+ {\cal O}(y,\nu(a),\nu(\hat a))
+ {\cal O}(\as^2)
\Bigg\}
\;.
\end{split}
\end{equation}
We see that the ratio of parton distribution functions disappears from ${\cal V}^{\rm pert}(t)$ if
\begin{equation}
\label{eq:GtoP}
P_{a\hat a}\!\left(z,\mu^2/z\right)
=
G_{a\hat a}\!\left(z,\mu^2/z\right)
\;.
\end{equation}
We compute $G$ directly from the splitting functions in the shower and this determines the evolution kernel $P$ for the parton distributions at finite values of $(1-z)$.

\section{Determining $\gamma_a(\bar\mu^2)$}
\label{sec:findingamma}

Eq.~(\ref{eq:GtoP}) gives us $P_{a\hat a}\!\left(z,\mu^2/z\right)$ at finite $(1-z)$ from the small $y$ limit $G_{a\hat a}$ of the initial state splitting functions in the shower. However the full splitting function is actually a distribution, with singular behavior at $z \to 1$, as indicated in eq.~(\ref{eq:evolution4}). We need to determine the constants $\gamma_a(\bar\mu^2)$ that appear in eq.~(\ref{eq:evolution4}). Essentially, these constants multiply $\delta(1-z)$ in the evolution kernel and are thus not present at finite $(1-z)$. However, we can determine the constants $\gamma_a(\bar\mu^2)$ from the momentum and flavor sum rules that guarantee that the total longitudinal momentum of the partons sums to the total longitudinal momentum of the proton and that the total flavor quantum numbers of the partons sums to the total flavor quantum numbers of the proton.

To proceed in a unified fashion, consider the quantity
\begin{equation*}
-\frac{d}{dt}\,
\sum_a c_a
\int_0^1\!d\eta_\La\
\eta_\La^N
\tilde f_{a/A}(\eta_{\La},\mu_\LA^2 e^{-t})
\;.
\end{equation*}
If we take
\begin{equation}
\begin{split}
c_a ={}& 1\hskip 1 cm {\rm for\ all}\ a
\;,
\\
N ={}& 1
\;,
\end{split}
\end{equation}
then the momentum sum rule implies that this quantity should be zero. If we let $q$ be a quark flavor and take
\begin{equation}
\begin{split}
c_a ={}& 
\begin{cases}
1 & a = q \\
-1 & a = \bar q \\
0 & {\rm otherwise}
\end{cases}
\;,
\\
N ={}& 0
\;,
\end{split}
\end{equation}
then the flavor sum rule for flavor $q$ implies that this quantity should be zero. Using eq.~(\ref{eq:evolution4}), we see that for either kind of sum rule
\begin{equation}
\begin{split}
0 ={}& -\frac{d}{dt}\,
\sum_a c_a
\int_0^1\!d\eta_\La\
\eta_\La^N
\tilde f_{a/A}(\eta_{\La},\mu_\LA^2 e^{-t})
\\
={}& \sum_{a,\hat a} c_a
\int_0^1\!d\eta_\La
\int_0^{1-}\!dz\
\eta_\La^N\,
\frac{\as(\eta_\La\mu_\LA^2 e^{-t}/z)}{2\pi}\,\frac{1}{z}
P_{a\hat a}\!\left(z,\eta_\La\mu_\LA^2 e^{-t}/z\right)\,
\tilde f_{\hat a/A}(\eta_{\La}/z, \mu_\LA^2 e^{-t})
\\& -
\sum_{a} c_a
\int_0^1\!d\eta_\La
\int_0^{1-}\!dz\
\eta_\La^N\,
\frac{\as(\eta_\La\mu_\LA^2 e^{-t})}{2\pi}
\left[
\frac{2C_a}{1-z}-\gamma_a(\eta_\La\mu_\LA^2 e^{-t})
\right]
\tilde f_{a/A}(\eta_{\La},\mu_\LA^2 e^{-t})
\\&
+ {\cal O}(\as^2)
\;.
\end{split}
\end{equation}
In the first term we change variables from $\eta_\La$ to $\hat \eta_\La = \eta_\La/z$, giving
\begin{equation}
\begin{split}
0 ={}& \sum_{a,\hat a} c_a
\int_0^{1-}\!dz
\int_0^1\!d\hat\eta_\La\
z^N \hat\eta_\La^N\,
\frac{\as(\hat\eta_\La \mu_\LA^2 e^{-t})}{2\pi}\,
P_{a\hat a}\!\left(z,\hat\eta_\La \mu_\LA^2 e^{-t}\right)\,
\tilde f_{\hat a/A}(\hat\eta_{\La}, \mu_\LA^2 e^{-t})
\\& -
\sum_{a} c_a
\int_0^{1-}\!\!dz
\int_0^1\!\!d\eta_\La\,
\eta_\La^N\,
\frac{\as(\eta_\La \mu_\LA^2 e^{-t})}{2\pi}\!
\left[
\frac{2C_a}{1-z}-\gamma_a(\eta_\La \mu_\LA^2 e^{-t})
\right]
\tilde f_{a/A}(\eta_{\La}, \mu_\LA^2 e^{-t})
\;.
\end{split}
\end{equation}
With a little manipulation, this is
\begin{equation}
\begin{split}
0 ={}& \sum_{\hat a}
\int_0^1\!d\eta_\La\
\eta_\La^N\,
\frac{\as(\eta_\La \mu_\LA^2 e^{-t})}{2\pi}\,
\tilde f_{\hat a/A}(\eta_{\La}, \mu_\LA^2 e^{-t})
\\ & \times
\sum_{a} c_a
\left\{
\int_0^{1-}\!dz
\left[
z^N\,
P_{a\hat a}\!\left(z,\eta_\La \mu_\LA^2 e^{-t}\right)
-\delta_{a \hat a}\, \frac{2C_{a}}{1-z}
\right]
+ \delta_{a \hat a} \gamma_{\hat a}(\eta_\La \mu_\LA^2 e^{-t})
\right\}
\;.
\end{split}
\end{equation}
The coefficient of $\tilde f_{\hat a/A}(\eta_{\La}, \mu_\LA^2 e^{-t})$ must vanish. Thus we need (setting $\eta_\La \mu_\LA^2 e^{-t} = \bar \mu^2$)
\begin{equation}
\begin{split}
c_{\hat a}\gamma_{\hat a}(\bar\mu^2) ={}& 
-\sum_{a} c_a
\int_0^{1-}\!dz
\left[
z^N\,
P_{a\hat a}\!\left(z,\bar\mu^2\right)
-\delta_{a \hat a}\, \frac{2C_{a}}{1-z}
\right]
\;.
\end{split}
\end{equation}
We now write this out in detail. The only nonzero functions $P_{\hat a a}$ are those for which there is a first order splitting graph for $\hat a \to a + f$ for some flavor $f$. Thus for any quark or antiquark flavors $q'$ and $q$, $P_{q' q} = 0$ unless $q' = q$. 

Let us examine the flavor sum rule for flavor $q$. Taking $\hat a = q$, we have
\begin{equation}
\begin{split}
\label{eq:qflavorforq}
\gamma_{q}(\bar\mu^2) ={}& 
-\int_0^{1-}\!dz
\left[
P_{qq}\!\left(z,\bar\mu^2\right)
- \frac{2 C_\LF}{1-z}
\right]
\;.
\end{split}
\end{equation}
Taking $\hat a = \Lg$ we have
\begin{equation}
\begin{split}
\label{eq:qflavorforg}
0 ={}& 
\int_0^{1-}\!dz
\left[
P_{q\Lg}\!\left(z,\bar\mu^2\right)
- P_{\bar q\Lg}\!\left(z,\bar\mu^2\right)
\right]
\;.
\end{split}
\end{equation}
Taking $\hat a = q'$ for any other flavor, we have simply 0 = 0. Now, charge conjugation invariance for the splitting functions dictates that
\begin{equation}
P_{q\Lg}\!\left(z,\bar\mu^2\right)
= P_{\bar q\Lg}\!\left(z,\bar\mu^2\right)
\;.
\end{equation}
Thus eq.~(\ref{eq:qflavorforg}) is automatically satisfied. This leaves eq.~(\ref{eq:qflavorforq}), which determines $\gamma_q$.

Now let us examine the momentum sum rule. Taking $\hat a = \Lg$, we have
\begin{equation}
\begin{split}
\label{eq:mommentumforg}
\gamma_{g}(\bar\mu^2) ={}& 
-\int_0^{1-}\!dz
\left[
z\,P_{\Lg\Lg}\!\left(z,\bar\mu^2\right)
- \frac{2 C_\LA}{1-z}
\right]
- 2\sum_{q\in {\cal Q}} \int_0^{1-}\!dz\
z\,P_{q\Lg}\!\left(z,\bar\mu^2\right)
\;.
\end{split}
\end{equation}
Here we sum over quark flavors ${\cal Q} = \{u,d,c,s,b\}$, not antiquark flavors, and then multiply the quark term by 2. Taking $\hat a$ to be a quark or antiquark flavor $q$, we have
\begin{equation}
\begin{split}
\label{eq:mommentumforq}
\gamma_{q}(\bar\mu^2) ={}& 
-\int_0^{1-}\!dz
\left[
z\,P_{qq}\!\left(z,\bar\mu^2\right)
- \frac{2 C_\LF}{1-z}
\right]
- \int_0^{1-}\!dz\
z\,P_{\Lg q}\!\left(z,\bar\mu^2\right)
\;.
\end{split}
\end{equation}
Now, eq.~(\ref{eq:mommentumforg}) determines $\gamma_\Lg$. Then eq.~(\ref{eq:mommentumforq}) would determine $\gamma_q$ except that we have already determined $\gamma_q$ in eq.~(\ref{eq:qflavorforq}). For these equations to be consistent, we need
\begin{equation}
\begin{split}
\label{eq:consistencyforq}
0 ={}& 
\int_0^{1-}\!dz
\left[
z\,P_{qq}\!\left(z,\bar\mu^2\right)
- P_{qq}\!\left(z,\bar\mu^2\right)
\right]
+ \int_0^{1-}\!dz\
z\,P_{\Lg q}\!\left(z,\bar\mu^2\right)
\;.
\end{split}
\end{equation}
Changing variables from $z$ to $1-z$ in the second integral, this is
\begin{equation}
\begin{split}
\label{eq:consistencyforq2}
0 ={}& 
\int_0^{1-}\!dz\
(1-z)
\left[
- P_{qq}\!\left(z,\bar\mu^2\right)
+ 
P_{\Lg q}\!\left(1-z,\bar\mu^2\right)
\right]
\;.
\end{split}
\end{equation}
The two functions $P_{qq}$ and $P_{\Lg q}$ both describe the splitting $q \to q + \Lg$ and differ by whether it is the quark or gluon that goes on to the hard interaction. We will find that these two functions are related by
\begin{equation}
\label{eq:PqqtoPgq}
P_{qq}\!\left(z,\bar\mu^2\right)
= 
P_{\Lg q}\!\left(1-z,\bar\mu^2\right)
\;.
\end{equation}
Because of this relation, the two formulas for calculating $\gamma_q$ give the same result.

\section{The result}
\label{sec:result}

We compute the small $y$ limit $G$ of the shower splitting functions and set the parton evolution kernels $P$ for finite $(1-z)$ equal to $G$ according to eq.~(\ref{eq:GtoP}). This gives
\begin{equation}
\begin{split}
\label{eq:kernels}
P_{qq}(z, \mu^2/z) = {}& C_{\rm F}\left[\frac{2}{(1-z)_{+}}-(1+z)
- 2z\,\frac{m(q)^2}{\mu^2}\right] 
\Theta\!\left((1-z) m(q)^2 < \mu^2\right)
\\&+ \gamma_{q}(\mu^2) \,\delta(1-z)
\;,\\
P_{\rm gg}(z, \mu^2/z) = {}& 2C_{\rm A}\left[\frac{1}{(1-z)_{+}} - 1 
+ \frac{1-z}{z}+ z(1-z)\right]
+ \gamma_{\rm g}(\mu^2)\,\delta(1-z)
\;,\\
P_{q{\rm g}}(z, \mu^2/z) = {}& T_\LR\left[1 - 2\,z\,\left(1-z\right) 
+ 2z\,\frac{m(q)^2}{\mu^2}\right]
\Theta\!\left(m(q)^2 < (1-z) \mu^2\right)
\;,
\\
P_{{\rm g}q}(z, \mu^2/z) = {}& C_{\rm F}\left[\frac{1+(1-z)^{2}}{z}
- 2z\,\frac{m(q)^2}{\mu^2}\right]
\Theta\!\left(z^2 m(q)^2 < (1-z)\mu^2\right)
\;.
\end{split}
\end{equation}
The constants $\gamma_a(\bar\mu^2)$ are computed according to Eqs.~(\ref{eq:mommentumforg}) and (\ref{eq:qflavorforq}), with the result
\begin{equation}
\begin{split}
\label{eq:gammas}
\gamma_\Lg(\bar\mu^2) ={}&
\frac{11}{6} C_{A} 
-  \frac{2T_{R}}{3} \sum_{q}
\sqrt{1 - \frac{4 m(q)^2}{\mu^2}}\,
\left(1 + \frac{2 m(q)^2}{\mu^2}\right)\,
\Theta\!\left({4m(q)^2} < {\mu^2}\right)
\;,
\\
\gamma_{q}(\bar\mu^2) ={}&
C_\LF\bigg\{
\frac32  
+ 2\log\left(1 + \frac{m(q)^2}{\mu^2}\right)
+ \frac{m(q)^2}{\mu^2}
\frac{
(2 +  {m(q)^2}/{\mu^2})}
{2(1 +  {m(q)^2}/{\mu^2})^2}
\bigg\}
\;.
\end{split}
\end{equation}
Note that the relation (\ref{eq:PqqtoPgq}) that allows a consistent calculation of $\gamma_q(\mu^2)$ does indeed hold. 

The theta functions that provide a lower limit on $\mu^2$ for a given $z$ in eq.~(\ref{eq:kernels}) are easy to understand. For $P_{qq}$ we consider a splitting of a quark with momentum $\hat p_\La = (\hat p_\La^+, \hat p_\La^-, \hat {\bm p}_\La)$ given by
\begin{equation}
\hat p_\La = \left(\frac{1}{z}\,p_\La^+,\, z\,\frac{m(q)^2}{2 p_\La^+},\, \bm 0\right)
\;.
\end{equation}
A daughter gluon is emitted into the final state with momentum
\begin{equation}
\hat p_{m+1} = \left(\frac{1-z}{z}\,p_\La^+,\, \frac{z}{1-z}\frac{\bm k^2}{2 p_\La^+},\, \bm k\right)
\;.
\end{equation}
This leaves a daughter quark heading toward the hard interaction carrying momentum $\hat p_\La -\hat p_{m+1}$. The daughter quark has virtuality $\mu^2 = -  (\hat p_\La -\hat p_{m+1})^2 + m(q)^2$  given by
\begin{equation}
\mu^2 = (1-z) m(q)^2 + \frac{1}{1-z}\,\bm k^2
\;.
\end{equation}
The minimum virtuality occurs when the transverse momentum $\bm k$ vanishes and we find $(1-z) m(q)^2 < \mu^2$, as in the first equation in eq.~(\ref{eq:kernels}). The other cases follow similarly.

The momentum sum rule constant $\gamma_\Lg(\mu^2)$ is of special interest. When $\mu^2$ is very large, each of $N_\Lf$ flavors of quark contributes and we have
\begin{equation}
\begin{split}
\gamma_\Lg(\mu^2) ={}&
\frac{11}{6} C_{A} 
-  \frac{2T_{R}}{3} N_\Lf
\;.
\end{split}
\end{equation}
However, when $\mu^2$ decreases to close to $4 m(q)^2$ for some flavor of quark, the contribution of that flavor begins to turn off because the splittings $g \to q + \bar q$ turn off. For $\mu^2 < {4m(q)^2}$, the contribution from quark $q$ turns off entirely.

\section{Difference between pdf's with and without mass}
\label{sec:differencemassmakes}

The parton distribution functions $f_{a/A}(\eta_{\La},\mu^{2})$ evolve according to eqs.~(\ref{eq:evolution1}),  (\ref{eq:kernels}), and (\ref{eq:gammas}). The $\MSbar$ parton distribution functions evolve according to the same equation with all of the quark masses set to zero, but with boundary conditions that set the quark distributions for heavy quarks to zero for $\mu^2 < m^2$, as in eq.~(\ref{eq:MSbarkernel}). For the purposes of this section, let us choose a modification of the $\MSbar$ scheme in which the boundary condition is at $\mu^2 = \lambda m^2$ for some $\lambda$ that is possibly not 1. We can call this the $\MSbar \lambda$ prescription. Thus the effective $g \to q$ evolution kernel is
\begin{equation}
\label{eq:MSbarkernelmod}
P_{qg}^{\overline{\rm MS}\lambda}(z,\mu^2/z)
=
T_\LR[1 - 2 z(1-z)]\,\Theta(\mu^2 > \lambda m^2)
\;.
\end{equation}
Let us work in a five flavor theory with the charm and bottom quark masses non-zero, while other quark masses are set to zero. Let us suppose that we set parton distribution functions $f_{a/A}(\eta_{\La},\mu^{2})$ equal to the $\MSbar \lambda$ parton distribution functions when the scale is smaller than the charm mass squared:
\begin{equation}
f_{a/A}(\eta_{\La},\mu^{2}) = f_{a/A}^{\MSbar\lambda}(\eta_{\La},\mu^{2})
\;,\hskip 1 cm
\mu^{2} < \lambda m(\Lc)^2
\;.
\end{equation}
Define the differences
\begin{equation}
\Delta f_{a/A}(\eta_{\La},\mu^{2}) = 
f_{a/A}(\eta_{\La},\mu^{2})
- f_{a/A}^{\MSbar\lambda}(\eta_{\La},\mu^{2})
\;.
\end{equation}
It is of interest to calculate the order $\as$ contribution to these differences.

Consider, for example, the change in the bottom quark distribution. Evidently
\begin{equation}
\begin{split}
\label{eq:DeltaEvolutionb}
\frac{d}{d\log(\mu^2)}\,
\Delta f_{\Lb/A}(\eta_{\La},\mu^{2})
={}& 
\sum_{\hat a} 
\int\!\frac{dz}{z}\
\frac{\as(\mu^2)}{2\pi} 
\Delta P_{\Lb\hat a}(z,\mu^2)\
f^{\MSbar\lambda}_{\hat a/A}(\eta_{\La}/z,\mu^{2})
+ {\cal O}(\as^2)
\;,
\end{split}
\end{equation}
where
\begin{equation}
\Delta P_{\Lb\hat a}(z,\mu^2)
= P_{\Lb\hat a}(z,\mu^2/z) - P^{\MSbar\lambda}_{\Lb\hat a}(z,\mu^2)
\;.
\end{equation}

We first note that the contribution from $\hat a = \Lb$ can be neglected because $f^{\MSbar\lambda}_{\Lb/A}(\eta_{\La}/z,\mu^{2})$ is nonzero only for $\mu^{2} > \lambda m(\Lb)^2 $ and the kernel is significantly nonzero only for $\mu^{2} \sim m(\Lb)^2$. In this region $f^{\MSbar\lambda}_{\Lb/A}(\eta_{\La}/z,\mu^{2})$ is itself of order $\as$,  so that $\hat a = \Lb$ contribution to  eq.~(\ref{eq:DeltaEvolutionb}) is of order $\as^2$. 

This leaves the contribution from $\hat a = \Lg$. If we integrate the differential equation, we have
\begin{equation}
\begin{split}
\label{eq:DeltaEvolutionb2}
\Delta f_{\Lb/A}(\eta_{\La},\mu^{2})
={}&
\int\!\frac{dz}{z}
\int_0^{\mu^2}\!\frac{d\bar \mu^2}{\bar \mu^2}
\frac{\as(\bar \mu^2)}{2\pi} \
\Delta P_{\Lb \Lg}(z,\bar\mu^2/z)\
f^{\MSbar\lambda}_{\Lg/A}(\eta_{\La}/z,\bar \mu^2)
+ {\cal O}(\as^2)
\;.
\end{split}
\end{equation}
Because of the structure of $\Delta P_{\Lb \Lg}(z,\bar\mu^2/z)$, the most important contributions for large $\mu^2$ to the integration over $\bar \mu^2$ come from $\bar \mu^2$ somewhere around $m(\Lb)^2$. A reasonable estimate of the most important integration region is $\bar \mu^2 \sim 4 m(\Lb)^2$. Thus at order $\as$ we can set $\mu^{2} = 4 m(\Lb)^2$ in the argument of $f^{\MSbar\lambda}_{\hat a/A}(\eta_{\La}/z,\mu^{2})$ and $\as(\mu^2)$. This gives
\begin{equation}
\begin{split}
\label{eq:DeltaEvolutionb3}
\Delta f_{\Lb/A}(\eta_{\La},\mu^{2})
={}&
\int\!\frac{dz}{z}\
\frac{\as(4 m(\Lb)^2)}{2\pi} 
\Delta R_{\Lb\Lg}(z,\mu^2)\
f^{\MSbar\lambda}_{\Lg/A}\!\left(
{\eta_{\La}}/{z},4 m(\Lb)^2
\right)
+ {\cal O}(\as^2)
\;,
\end{split}
\end{equation}
where
\begin{equation}
\label{eq:DeltaR}
\Delta R_{\Lb\Lg}(z,\mu^2)
=
\int_0^{\mu^2}\!\frac{d\bar\mu^2}{\bar\mu^2}\
\Delta P_{\Lb \Lg}(z,\bar\mu^2)
\;.
\end{equation}

We learn three things. First, $\Delta f_{\Lb/A}(\eta_{\La},\mu^{2}) = 0$ for $\mu^2 < \min(1,\lambda)\, m(\Lb)^2$ because both versions of the parton distribution for b quarks vanish there. Second, $\Delta f_{\Lb/A}(\eta_{\La},\mu^{2})$ changes as $\mu^2$ increases; if $\lambda = 1$, it becomes negative because the splittings $\Lg \to \Lb + \bar{\Lb}$ turn on more slowly with a physical treatment of the threshold than with the $\MSbar$ treatment. Third, for very large $\mu^2$, the difference stays finite because the integral in eq.~(\ref{eq:DeltaR}) is finite in the limit $\mu^2 \to \infty$. In fact
\begin{equation}
\label{eq:DeltaR2}
\Delta R_{\Lb\Lg}(z,\infty)
=
T_\LR\big\{
[1-2 z(1-z)]\log(\lambda(1-z)) + 2 z (1-z)
\big\}
\;.
\end{equation}
This function is negative with a logarithmic singularity for $z \to 1$. For small $z$ it is positive. If we choose the standard $\MSbar$ prescription $\lambda = 1$, then the relevant convolution with the gluon distribution is negative, so that there are fewer bottom quarks with the shower evolution of the partons than with $\MSbar$ evolution, as we will see in section~\ref{sec:pdfbehavior}. Increasing $\lambda$ makes the difference with the $\MSbar \lambda$ bottom quark distribution smaller.

Evidently, analogous results apply for the charm quark distribution.

For the gluon distribution, similar reasoning gives
\begin{equation}
\begin{split}
\label{eq:DeltaEvolutiong3}
\Delta f_{\Lg/A}(\eta_{\La},\mu^{2})
={}&
\int\!\frac{dz}{z}\
\frac{\as(4 m(\Lb)^2)}{2\pi} 
\Delta R_{\Lg\Lg}(z,\mu^2)\
f^{\MSbar\lambda}_{\Lg/A}\!\left(
{\eta_{\La}}/{z},4 m(\Lb)^2
\right)
+ {\cal O}(\as^2)
\;,
\end{split}
\end{equation}
where
\begin{equation}
\label{eq:DeltaRg}
\Delta R_{\Lg\Lg}(z,\mu^2)
=
\int_0^{\mu^2}\!\frac{d\bar\mu^2}{\bar\mu^2}\
\Delta P_{\Lg \Lg}(z,\bar\mu^2)
\;.
\end{equation}
The evolution kernel $P_{\Lg \Lg}$ is the same as the $\MSbar\lambda$ version except for the term $\gamma_\Lg(\mu^2)\,\delta(1-z)$. Thus
\begin{equation}
\label{eq:DeltaRg2}
\Delta R_{\Lg\Lg}(z,\mu^2)
=
\delta(1-z)
\int_0^{\mu^2}\!\frac{d\bar\mu^2}{\bar\mu^2}\
\left[
\gamma_\Lg(\mu^2)
- \frac{11}{6}\,C_\LA + \frac{2 T_\LR}{3}\sum_q \Theta(\lambda m(q)^2 < \mu^2)
\right]
\;.
\end{equation}
After performing the integration, the result for large $\mu^2$ is very simple
\begin{equation}
\label{eq:DeltaRg3}
\Delta R_{\Lg\Lg}(z,\infty)
=
\delta(1-z)\,
\frac{10 T_\LR}{9}\,\left(1-\frac{3}{5}\log(\lambda)\right)\sum_q \Theta(0 < m(q)^2)
\;.
\end{equation}
The sum simply counts the number of quarks treated as massive, which is normally 2. The coefficient of $\delta(1-z)$ is generally not large. It is positive for $\lambda = 1$ and vanishes when $\lambda = e^{5/3} \approx 
5.3$.

\section{Behavior of the parton distributions}
\label{sec:pdfbehavior}

The parton distribution functions introduced in this paper have a different definition from the conventional $\MSbar$ parton distributions. Thus one should fit them to data using perturbation theory for deeply inelastic lepton scattering and other hard scattering processes that help to determine parton distributions. Needless to say, this is a very big project and we have not attempted it. However, parton showers are, at least at present, accurate only to lowest order in QCD perturbation theory. At this order, we may hope that the following scheme suffices. We take a standard set of $\MSbar$ parton distributions. For this paper, we have used the MSTW 2008 leading order central fit \cite{MSTW}.\footnote{In refs.~\cite{deductor} and \cite{ShowerTime}, we use a different set.} These are defined by applying ordinary $\MSbar$ evolution to the parton distributions at a starting scale $Q_{\rm fit}$. For the MSTW 2008 set, the starting scale is $Q_{\rm fit} = 1 \GeV$. Instead, we can define shower parton distributions by applying the evolution equation~(\ref{eq:evolution1}) to the parton distributions at the starting scale $Q_{\rm fit}$. In this section, the parton distributions thus defined are labelled simply as ``shower.'' We also display distributions labelled as $\MSbar$, which are defined by applying the standard $\MSbar$ lowest order evolution to the parton distributions at the starting scale $Q_{\rm fit}$. This is the same as the MSTW 2008 LO set. Finally, we display distributions labelled as $\MSbar \lambda$, which are defined by letting the partons evolve from the starting scale $Q_{\rm fit}$ using the standard $\MSbar$ lowest order evolution kernels but with the boundary condition that heavy quark evolution (for c and b quarks) starts at $\mu^2 = \lambda m^2$, as discussed in the previous section. This amounts to redefining the renormalization prescription for the heavy quark distribution functions so that one subtracts not only an ultraviolet pole term and a conventional finite term proportional to $(\log(4\pi) - \gamma_{\rm E})$, but also a finite term proportional to $\log(\lambda)$. Naturally, this would entail a corresponding change in the factorization subtraction for next-to-leading order hard scattering graphs. In this section, we choose $\lambda = 4$, so that the heavy quark threshold is at $\mu^2 = 4 m^2$.

\begin{figure}
\centerline{\includegraphics[height = 7 cm]{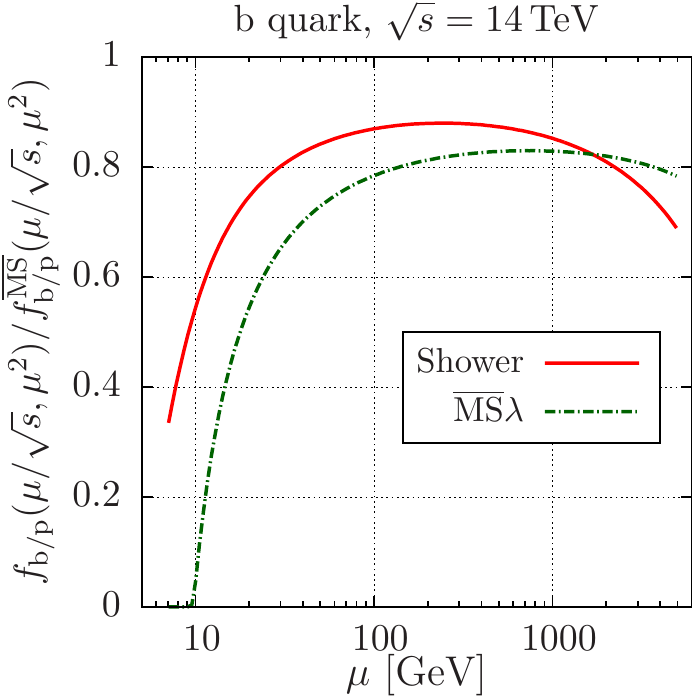}}
\caption{Ratio of the shower b-quark distribution to the $\MSbar$ b-quark distribution as it applies to the hard scattering at scale $\mu$. Also shown is the ratio of the $\MSbar \lambda$ b-quark distribution with $\lambda = 4$ to the standard $\MSbar$ b-quark distribution.}
\label{fig:pdfratio}
\end{figure}

Consider first what happens at the hard scattering that serves as the starting point for parton showers. Suppose that the hard scattering has scale $\mu^2 = \hat s$. Then the partons that produce the hard scattering have momentum fractions $\eta$ given by $\hat s = \eta_\La \eta_\Lb s$. Assume that the hard scattering is at central rapidity, so that $\eta_\La \approx \eta_\Lb$. Then $\eta_\La \approx \eta_\Lb \approx \mu/\sqrt{s}$. Thus we use parton distribution functions $f_{a/A}(\mu/\sqrt{s},\mu^2)$. We take $\sqrt s = 14\ \text{TeV}$. In figure \ref{fig:pdfratio}, we plot the ratio of $f_{a/A}(\mu/\sqrt{s},\mu^2)$ for b quarks to the corresponding b-quark distribution function in the $\MSbar$ prescription. We see that in the interesting range $100 \GeV < \mu < 2000 \GeV$ this ratio is around 0.8. The reason, of course, is that physical b-quark evolution starts more slowly than $\MSbar$ evolution. This is a perturbative effect, as analyzed in the previous section. We show also the ratio of $f_{\Lb/A}(\mu/\sqrt{s},\mu^2)$ in the $\MSbar \lambda$ prescription with $\lambda = 4$ to the b-quark distribution function in the standard ($\lambda = 1$) $\MSbar$ prescription. This ratio is also around 0.8.

\begin{figure}
\begin{center}
\begin{tabular}{@{\hspace{-0.3mm}}r@{\hspace{4.5mm}}r}
\includegraphics[height = 7 cm]{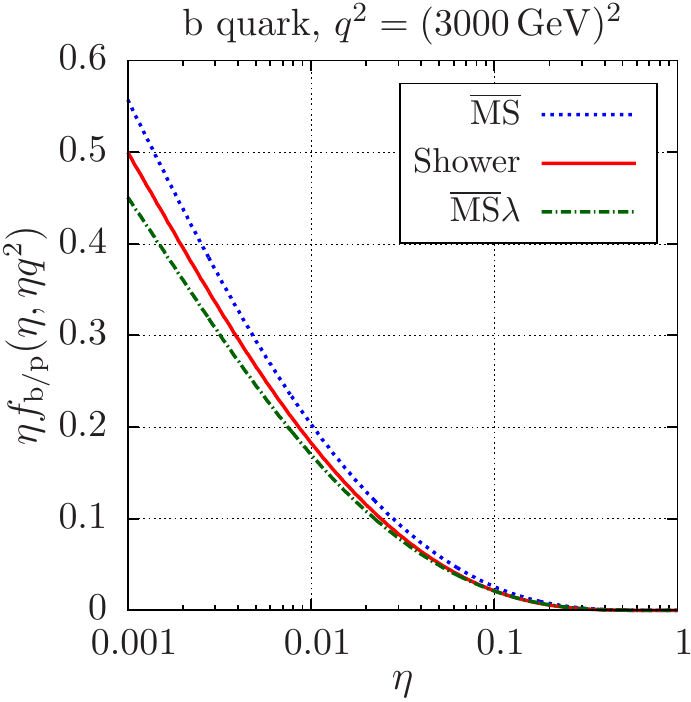} &
\includegraphics[height = 7 cm]{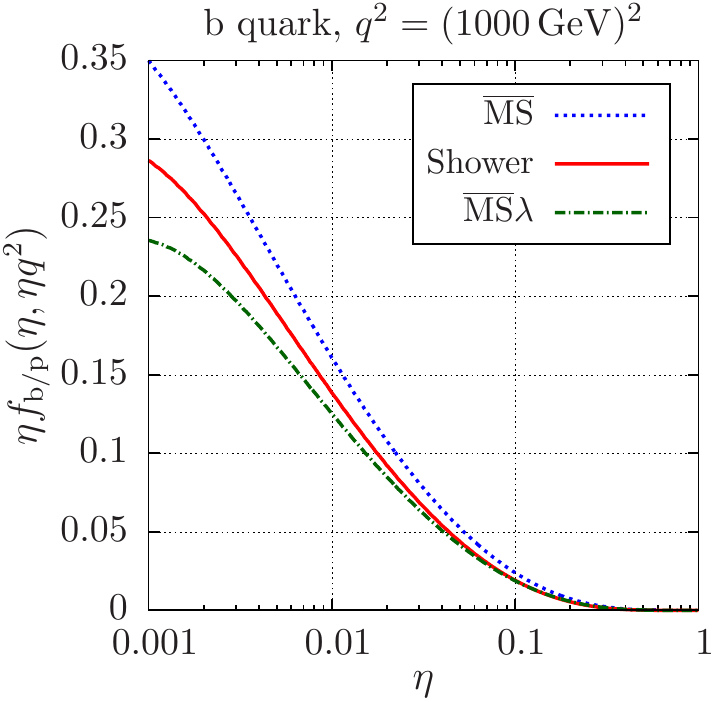}\\
~&~\\
\includegraphics[height = 7 cm]{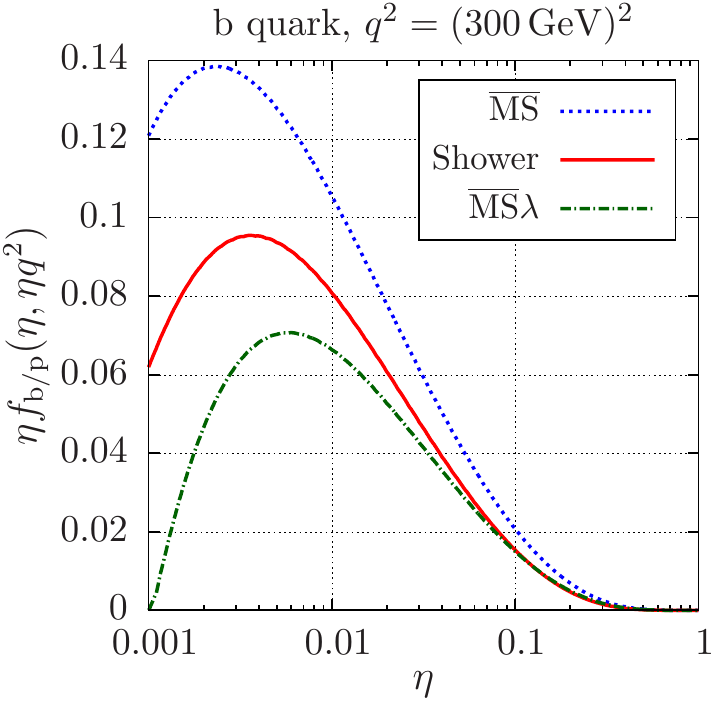} &
\includegraphics[height = 7 cm]{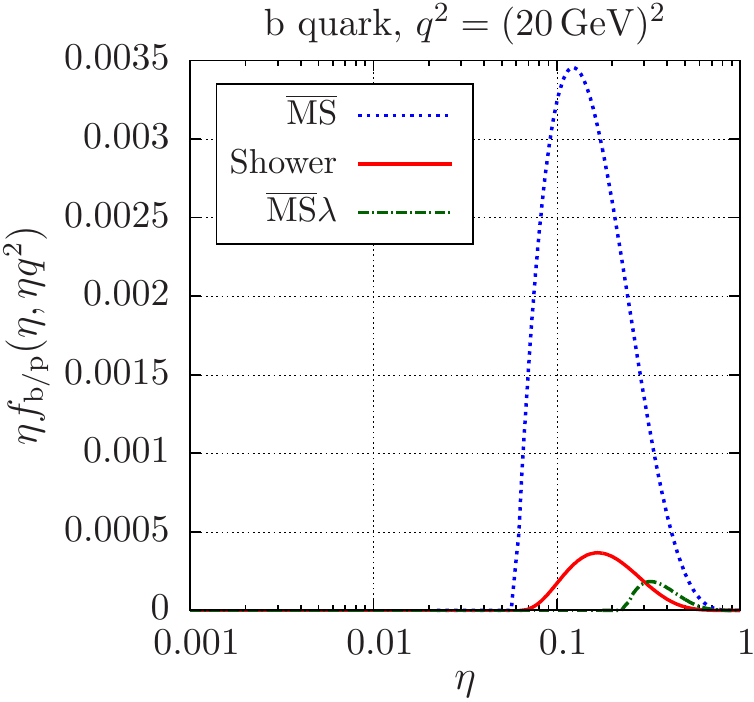}
\end{tabular}
\end{center}
\caption{Development of the b-quark distribution with increasing scale parameter. In each plot we show the $\eta$ dependence at fixed shower time $t$, so that $q^2 = \mu_\LA^2 e^{-t}$ is fixed.
There are curves for shower, $\MSbar$, and $\MSbar \lambda$ parton distributions for $\lambda = 4$.}
\label{fig:bquarkplots}
\end{figure}

Now we look at the parton distributions from the point of view of the shower. We consider $\tilde f_{a/A}(\eta, q^2)$ as a function of the momentum fraction $\eta$ at fixed shower time $t$, with $q^2 = \mu_\LA^2 e^{-t}$. These functions are related to the functions $f_{a/A}(\eta, \mu^2)$ by eq.~(\ref{eq:pdfrelationLO}), which gives
\begin{equation}
\tilde f_{a/A}(\eta, q^2) = f_{a/A}(\eta, \eta\, q^2)
\;.
\end{equation}
In figure \ref{fig:bquarkplots}, we plot the b-quark distribution in a proton, $f_{\Lb/\Lp}(\eta, \eta\, q^2)$, as functions of $\eta$ at fixed $q^2$. If we imagine starting at a hard interaction at central rapidity with a scale $Q_0^2 \approx (640 \GeV)^2$, then $\eta \approx \sqrt{Q_0^2/s} \approx 0.05$. With $\eta q^2 = Q_0^2$ we have $q \approx 3000 \GeV$. In the first panel of figure \ref{fig:bquarkplots}, we show $f_{\Lb/\Lp}(\eta, \eta\, q^2)$ versus $\eta$ at $q = 3000 \GeV$. We also show the b-quark distributions in the $\MSbar$ convention and in the $\MSbar \lambda$ convention with $\lambda = 4$. Now with ``backward evolution'' for the initial state, we move to smaller $q^2$ and larger $\eta$.  In the second panel of figure \ref{fig:bquarkplots}, we show the b-quark distribution versus $\eta$ at $q = 1000 \GeV$. The value of the b-quark distribution has started to decrease, which means that shower evolution will often turn a b quark into an incoming gluon. In the third panel, we show the b-quark distribution versus $\eta$ at $q = 300 \GeV$. The value of the b-quark distribution has now decreased dramatically: a substantial fraction of the b quarks have been turned into incoming gluons. Finally, in the fourth panel, we show the b-quark distribution versus $\eta$ at $q = 20 \GeV$. This is very close to the threshold. All but a small fraction of the b quarks have disappeared and only a limited range of $\eta$ is allowed for those that remain. With $\MSbar$ evolution, many more b quarks would remain. That is, the discrepancy  is substantial between evolution that follows the Feynman diagrams for $\Lg \to \Lb + \bar \Lb$ with $m_\Lb > 0$ and $\MSbar$ evolution.

\begin{figure}
\centerline{\includegraphics[height = 7 cm]{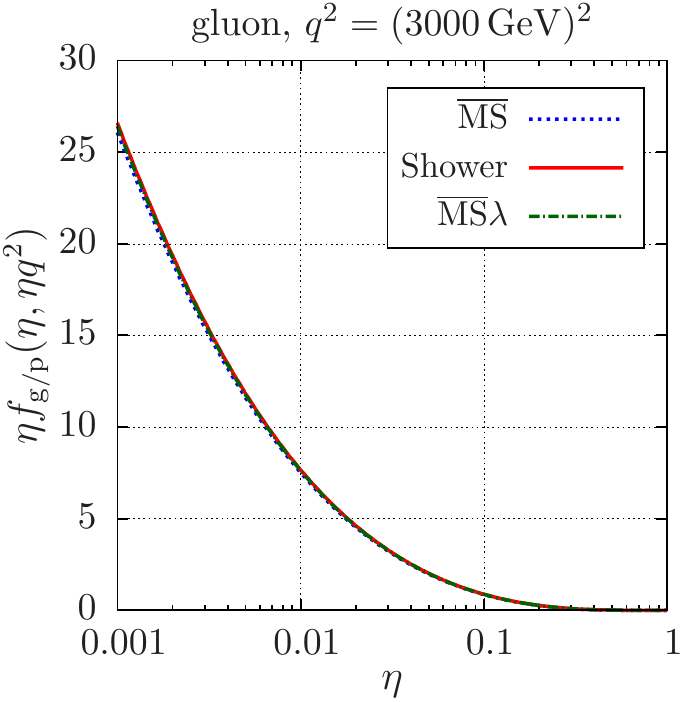}}
\caption{Dependence of the gluon distribution on $\eta$ at a large value $q = 3000 \GeV$ of the shower evolution scale parameter. There are curves for shower, $\MSbar$, and $\MSbar \lambda$ parton distributions for $\lambda = 4$. However, the differences are small.}
\label{fig:gluon10000}
\end{figure}

One may wonder what happens to the gluon distribution. In figure \ref{fig:gluon10000}, we show the distribution $f_{\Lg/\Lp}(\eta, \eta\, q^2)$ for gluons at fixed $q^2 = (3000 \GeV)^2$ for shower, $\MSbar$, and $\MSbar \lambda$ parton distributions for $\lambda = 4$. We note that there is hardly any difference.

\section{A small modification}
\label{sec:NLOaddition}

If we were to add one more order of perturbation theory to our parton evolution, we would have the evolution equation
\begin{equation}
\begin{split}
\label{eq:evolution4bis}
\frac{d\,\tilde f_{a/A}(\eta_{\La},\mu^{2}/\eta_\La)}{d\log(\mu^2)}
={}&
\sum_{\hat a} 
\int\!\frac{dz}{z}\
\frac{\as(\mu^2/z)}{2\pi}
P_{a\hat a}\!\left(z,\mu^2/z\right)\,
\tilde f_{\hat a/A}(\eta_{\La}/z, \mu^2/\eta_\La)
\\ &+
\sum_{\hat a} 
\int\!\frac{dz}{z}\
\left[\frac{\as(\mu^2/z)}{2\pi}\right]^2
P_{a\hat a}^{(2)}\!\left(z,\mu^2/z\right)\,
\tilde f_{\hat a/A}(\eta_{\La}/z, \mu^2/\eta_\La)
\;.
\end{split}
\end{equation}
In our analysis, we have regularly dropped contributions to $P_{a\hat a}^{(2)}$, since we have only a leading order shower. However, we find it helpful to use a limited version of $P_{a\hat a}^{(2)}$ in our parton evolution:
\begin{equation}
\label{eq:NLOkernel}
P_{a\hat a}^{(2)}(z,\mu^2/z) = - 2 \pi \beta_0 \log[\lambda_\LR]\, P_{a\hat a}\!\left(z,\mu^2/z\right)
\;,
\end{equation}
where $\beta_0 = (33 - 2 n_\Lf)/(12 \pi)$ is the first coefficient in the QCD $\beta$ function and where \cite{lambdaR}
\begin{equation}
\label{eq:lambdaR}
\lambda_\LR = \exp\left(
- \frac{C_\LA (67 - 3 \pi^2) - 10 n_\Lf}{3(33 - 2 n_\Lf)}
\right)
\approx 0.4
\;.
\end{equation}
The terms in eq.~(\ref{eq:NLOkernel}) proportional to $1/(1-z)$ appear in the exact $\MSbar$ parton evolution kernel at next-to-leading order. This modification of the evolution kernel amounts to changing $\mu^2$ in the argument of $\as$ in the evolution equation to $\lambda_\LR\, \mu^2$. 

In fact, we do include a factor $\lambda_\LR$ in the argument of $\as$ in shower evolution, as discussed in section \ref{sec:alphas}. Accordingly, we also use $\as(\lambda_\LR \mu^2)$ in place of $\as(\mu^2)$ in the evolution equation for the parton distributions. Thus we effectively include the term given in eq.~(\ref{eq:NLOkernel}) in the evolution of the parton distributions used in \textsc{Deductor}. However, we have not used this modification in the comparisons of massive and massless evolution presented in the sections \ref{sec:differencemassmakes} and \ref{sec:pdfbehavior}.

\section{Conclusions}
\label{sec:conclusions}

When the initial state evolution of a parton shower is organized according to the standard prescription of ref.~\cite{Sjostrand:1985xi}, the probabilities for parton splittings involve ratios of parton distribution functions. We have argued that in order for this to be physically consistent, the kernels of the evolution equation for the parton distributions need to be consistent with the splitting functions in the shower. In the case that the initial state partons can have non-zero masses, as in \textsc{Deductor} \cite{deductor}, this means that the parton evolution kernels cannot be the standard $\MSbar$ kernels. 

In this paper, we have deduced what the revised parton evolution kernels should be in order to match the shower evolution in \textsc{Deductor} to first order in $\as$.

Numerical investigations presented in section~\ref{sec:differencemassmakes} show that the modification of the evolution strongly affects the distribution functions for heavy quarks at evolution scales comparable to the square of the heavy quark mass. This effect shrinks as the evolution scale increases. The gluon distribution function is not much affected at any scale.

There is work to be done to understand these issues better. We would like to see what happens, for instance, if we keep non-zero masses but use $k_\LT$ ordering for the shower evolution instead of the ordering specified in eq.~(\ref{eq:showertime}). We would also like to have an operator definition of the modified parton distribution functions, analogous to that for $\MSbar$ parton distribution functions \cite{CSpartons}.

\acknowledgments{ 
This work was supported in part by the United States Department of Energy and by the Helmoltz Alliance ``Physics at the Terascale." We thank Voica Radescu of the \textsc{HeraFitter} group for providing the parton distribution functions that we use in the \textsc{Deductor} code.
}
\appendix

\section{The splitting functions}
\label{sec:splittingfctns}

We have argued that the parton splitting functions $P_{a\hat a}\!\left(z,\mu^2/z\right)$ should be given by eq.~(\ref{eq:GtoP}), which equates these functions to functions $G_{a\hat a}\!\left(z,\mu^2/z\right)$ that are defined in eq.~(\ref{eq:gtoG}) to be the small virtuality limits of parton splitting functions $g_{a\hat a}\!\left(z,{\mu^2}/{z},\{p,f\}_{m}\right)$. In this appendix, we calculate the functions $G_{a\hat a}\!\left(z,\mu^2/z\right)$.

\subsection{The virtual splitting operator and the splitting functions}
\label{eq:Vstructure}

In order to find $G_{a\hat a}\!\left(z,\mu^2/z\right)$, we seek the splitting function $g_{a\hat a}$ that appears in eq.~(\ref{eq:Veigenvalue}), which we repeat here:
\begin{equation}
\begin{split}
\label{eq:Veigenvaluebis}
\lambda^{\cal V}_{\La\La}(\{p,f\}_{m},t)
={}&
\sum_{\hat a} 
\int_0^{1-}\!dz\
\frac{\as(\mu^2)}{2\pi}\,
\frac{1}{z}\,
g_{a\hat a}\!\left(z,{\mu^2}/{z},\{p,f\}_{m}\right)\,
\frac{\tilde f_{\hat a/A}(\eta_{\La}/z, \mu^2/\eta_\La)}
{\tilde f_{a/A}(\eta_\La,\mu^2/\eta_\La)}
\;.
\end{split}
\end{equation}
Here $\lambda^{\cal V}_{\La\La}$ is the eigenvalue of a part ${\cal V}_{\La\La}(t)$ of the virtual splitting operator, as defined in eq.~(\ref{eq:Veigenvaluedef}),
\begin{equation}
\label{eq:Veigenvaluedefbis}
{\cal V}_{\La\La}(t)\sket{\{p,f,s',c',s,c\}_{m}}
= \lambda^{\cal V}_{\La\La}(\{p,f\}_{m},t)\sket{\{p,f,s',c',s,c\}_{m}}
\;.
\end{equation}
In general, the virtual splitting operator ${\cal V}(t)$ is determined from the real splitting operator ${\cal H}_\LI(t)$ by  eq.~(3.47) of ref.~\cite{NSI},
\begin{equation}
0 = \sbra{1}[{\cal H}_\LI(t) - {\cal V}(t)]
\;,          
\end{equation}
where multiplication by $\sbra{1}$ represents making an inclusive measurement and the inner product of $\sbra{1}$ with a statistical basis state is

\begin{equation}
\sbrax{1}\sket{\{p,f,s',c',s,c\}_{m}}
= \brax{\{s'\}_{m}}\ket{\{s\}_{m}}
\brax{\{c'\}_{m}}\ket{\{c\}_{m}}
\;.
\end{equation}
Thus
\begin{equation}
\label{eq:findeigenvalue}
\brax{\{s'\}_{m}}\ket{\{s\}_{m}}
\brax{\{c'\}_{m}}\ket{\{c\}_{m}}
\lambda^{\cal V}_{\La\La}(\{p,f\}_{m},t)
=
\sbra{1}{\cal H}_{\LI,\La\La}(t)\sket{\{p,f,s',c',s,c\}_{m}}
\;.
\end{equation}
As explained in section~\ref{sec:findingP}, the subscript ``aa'' here means that we are to take the part of ${\cal H}_\LI(t)$ that describes splittings of incoming parton ``a'' and comes from graphs (in a physical gauge) in which parton ``a'' splits in both the quantum ket state and the quantum bra state. There are other contributions ${\cal V}_{\La k}$ that come from interference graphs. As explained in section~\ref{sec:findingP}, these are soft gluon contributions and do not contain a ratio of parton distribution functions, so we can ignore them.

To analyze eq.~(\ref{eq:findeigenvalue}), we begin with eqs.~(12.20) and (12.21) of ref.~\cite{NSI}. We construct ${\cal H}_{\LI,\La\La}(t)$ by keeping only the terms corresponding to ``aa'' graphs:
\begin{equation}
\begin{split}
\label{eq:inclusiveH}
\sbra{1}&{\cal H}_{\rm I,aa}(t)
\sket{\{p,f,s',c',s,c\}_{m}}
=
\\&
\sum_{\hat a}
\int d\zeta_{\rm p}\
\theta(\zeta_{\rm p} \in \varGamma_{\La}(\{p\}_{m},\zeta_{\rm f}))\,
\delta\!\left(
t - \log\left(\frac{\eta_\La\, \mu_\LA^2}
{|(\hat p_\La 
-
 \hat p_{m+1})^2 - m(a)^2|}\right)
\right)
\\&\times
\frac
{n_\Lc(a)\,\eta_{\La}}
{n_\Lc(\hat a)\,
 \hat \eta_{\La}}\,
\frac{
\tilde f_{\hat a/A}(\hat \eta_{\La},\mu_\LA^2 e^{-t})}
{\tilde f_{a/A}(\eta_{\La},\mu_\LA^2 e^{-t})}\,
\brax{\{s'\}_{m}}\ket{\{s\}_{m}}
\\&\times
\biggl\{
\theta(\hat a \ne a)\
\brax{\{c'\}_{m}}\ket{\{c\}_{m}}\,
T_{\rm R}\
\overline w_{\La\La}(\{\hat f,\hat p\}_{m+1})
\\ &\ \ -
\theta(\hat a = a)\sum_{k \ne \La}
\bra{\{c'\}_{m}}
g_{\La k}(\{\hat f\}_{m+1})
\ket{\{c\}_{m}}
\overline w_{\La\La}(\{\hat f,\hat p\}_{m+1})
\bigg\}
\;.
\end{split}
\end{equation}
This formula requires a bit of explanation. We examine the integration measure $d\zeta_{\rm p}$ in the following subsection. The delta function defines the shower time according to eq.~(\ref{eq:showertime}), which is different definition than that used in ref.~\cite{NSI}. The parton flux factor is described in eq.~(\ref{eq:HtoHpert}). The splitting functions $\overline w_{\La\La}$ are given in ref.~\cite{NSI}. The symbol $g_{\La k}$ represent an operator on the color space\footnote{We hope that these color operators $g_{\La k}(\{\hat f\}_{m+1})$, defined in ref.~{\cite{NSI}}, will not be confused with the functions $g_{\La k}\!\left(z, {\mu^2}/{z} ,\{p,f\}_{m}\right)^{c'c}_{\bar c'\bar c}$ and $g_{\La \La}\!\left(z, {\mu^2}/{z} ,\{p,f\}_{m}\right)$ used in section \ref{sec:findingP}.} that obeys the color identity
\begin{equation}
\sum_{k \ne \La}
\bra{\{c'\}_{m}}
g_{\La k}(\{\hat f\}_{m+1})
\ket{\{c\}_{m}}
= -
\bra{\{c'\}_{m}}
g_{\La \La}(\{\hat f\}_{m+1})
\ket{\{c\}_{m}}
\;.
\end{equation}
Furthermore,
\begin{equation}
\bra{\{c'\}_{m}}
g_{\La \La}(\{\hat f\}_{m+1})
\ket{\{c\}_{m}}
= \brax{\{c'\}_{m}}\ket{\{c\}_{m}}\
C(\hat a, a)
\;,
\end{equation}
where
\begin{equation}
C(\hat a, a) = 
\begin{cases}
C_\LF & (\hat a, a) = (q,q), (\bar q, \bar q), (g,q)
\text{\ or\ }(g,\bar q) \\
C_\LA & (\hat a, a) = (\Lg,\Lg) \\
T_\LR & (\hat a, a) = (q,\Lg) \text{\ or\ } (\bar q, \Lg)
\end{cases}
\;.
\end{equation}
Thus $\sbra{1}{\cal H}_{\rm I}(t) \sket{\{p,f,s',c',s,c\}_{m}}$ contains factors of $\brax{\{s'\}_{m}}\ket{\{s\}_{m}}$ and $\brax{\{c'\}_{m}}\ket{\{c\}_{m}}$, so that we can identify the eigenvalue $\lambda^{\cal V}_{\La\La}$ according to eq.~(\ref{eq:findeigenvalue}). We find
\begin{equation}
\begin{split}
\label{eq:lambdaV1}
\lambda^{\cal V}_{\La\La}&(\{p,f\}_{m},t)
=
\\&
\sum_{\hat a}
\int d\zeta_{\rm p}\
\theta(\zeta_{\rm p} \in \varGamma_{\La}(\{p\}_{m},\zeta_{\rm f}))\
\delta\!\left(
t - \log\left(\frac{\eta_\La\,\mu_\LA^2}
{|(\hat p_\La 
-
 \hat p_{m+1})^2 - m(a)^2|}\right)
\right)
\\&\times
\frac
{n_\Lc(a)\,\eta_{\La}}
{n_\Lc(\hat a)\,
 \hat \eta_{\La}}\,
\frac{
\tilde f_{\hat a/A}(\hat \eta_{\La},\mu_\LA^2 e^{-t})}
{\tilde f_{a/A}(\eta_{\La},\mu_\LA^2 e^{-t})}
\,
C(\hat a, a)\,
\overline w_{\La\La}(\{\hat f,\hat p\}_{m+1})
\;.
\end{split}
\end{equation}

\subsection{Initial state splitting kinematics}
\label{sec:kinematics}

In order to proceed, we need to specify in some detail the kinematics of an initial state splitting in our version of a parton shower, in which partons can have non-zero masses and in which we implement momentum conservation in a somewhat different way from other parton shower algorithms.

We consider a splitting of initial state parton ``a'' with momentum $p_\La$. In general, we denote the momentum of parton $i$ before the splitting by $p_i$ and after the splitting by $\hat p_i$. Before the splitting, there are $m$ final state partons. The splitting creates a new final state parton with momentum $\hat p_{m+1}$. Here ``before'' and ``after'' are in the sense of backward evolution, so that the initial state parton with momentum $\hat p_\La$ evolves going forward in physical time to the partons with momenta $p_\La$ and $\hat p_{m+1}$. The two initial state partons have momenta given by eq.~(\ref{eq:papbdef}), in which $\eta_\La$ and $\eta_\Lb$ are the respective momentum fractions and $p_\LA$ and $p_\LB$ are the initial hadron momenta modified slightly so that they are lightlike, with $2 p_\LA \cdot p_\LB = s$.

It will prove convenient to define lightlike vectors $n_\La$ and $n_\Lb$ by
\begin{equation}
\begin{split}
n_\La ={}& \eta_\La\, p_\LA
\;,\qquad
n_\Lb = \eta_\Lb\, p_\LB
\;.
\end{split}
\end{equation}
Then
\begin{equation}
2 n_\La \cdot n_\Lb = \eta_\La \eta_\Lb s
\;.
\end{equation}
We also define dimensionless mass squared variables by
\begin{equation}
\begin{split}
\label{eq:mudefs}
\nu_\La ={}& \frac{m(a)^2}{2 n_\La \cdot n_\Lb}
\;,\quad
\hat\nu_\La = \frac{m(\hat a)^2}{2 n_\La \cdot n_\Lb}
\;,\quad
\nu_\Lb = \frac{m(b)^2}{2 n_\La \cdot n_\Lb}
\;,\quad
\hat\nu_{m+1} ={} \frac{m(\hat f_{m+1})^2}{2 n_\La \cdot n_\Lb}
\;.
\end{split}
\end{equation}
This is a somewhat more compact version of the definition in eq.~(\ref{eq:nufdef}). With this notation, the incoming parton momenta are
\begin{equation}
\begin{split}
p_\La ={}& n_\La + \nu_\La\, n_\Lb
\;,
\\
p_\Lb ={}& n_\Lb + \nu_\Lb\, n_\La
\;.
\end{split}
\end{equation}

After the splitting, parton ``b'' remains the same,
\begin{equation}
\hat p_\Lb = p_\Lb
\;.
\end{equation}
However, parton ``a'' gets a new momentum,
\begin{equation}
\hat p_\La = \frac{\hat\eta_\La}{\eta_\La}\, n_\La 
+ \frac{\eta_\La}{\hat\eta_\La}\,\hat\nu_\La\, n_\Lb
\;.
\end{equation}
We define a momentum fraction for the splitting,
\begin{equation}
z = \frac{\eta_\La}{\hat \eta_\La}
\;.
\end{equation}
Then
\begin{equation}
\label{hatpLa}
\hat p_\La = \frac{1}{z}\, n_\La 
+z\,\hat\nu_\La\, n_\Lb
\;.
\end{equation}

We define a virtuality variable $y$ by
\begin{equation}
\label{eq:ydef}
(\hat p_{\La} - \hat p_{m+1})^2 - m(a)^2 = 
- y\,2 n_\La\cdot n_\Lb
\;.
\end{equation}
We can express this using $\hat p_\La \cdot \hat p_{m+1}$ as
\begin{equation}
\label{eq:ydef2}
2\hat p_\La \cdot \hat p_{m+1} = 
\left[
y  
+ \hat\nu_\La + \hat\nu_{m+1} - \nu_\La
\right]2 n_\La\cdot n_\Lb
\;.
\end{equation}
It is well to note here that we will later use approximations based on $y \ll 1$, $\hat\nu_\La \ll 1$, $\nu_\La \ll 1$, $\nu_\Lb \ll 1$, and $\hat\nu_{m+1} \ll 1$. However, we do not take $y$ to be either much larger or much smaller than the dimensionless mass variables. When shower evolution reaches a stage near to heavy quark thresholds, these variables are comparable.

At this point, we need to remind ourselves about a subtle point that affects shower kinematics. Before the splitting, we know $p_\La$. At the splitting, parton $m+1$ with momentum $\hat p_{m+1}$ is emitted. Since $\hat p_{m+1}$ has four components but $\hat p_{m+1}^2 = m(f_{m+1})^2$, the momentum $\hat p_{m+1}$ can be described using three splitting variables. Knowing $\hat p_{m+1}$ should then determine $\hat p_\La$. However, we cannot simply set $\hat p_\La$ to $p_\La + \hat p_{m+1}$ because then $\hat p_\La$ will not be on-shell and it will not have zero components transverse to the beam. Instead, we need to take a small amount of momentum from elsewhere in the event and supply it to $\hat p_\La$. The method chosen in ref.~\cite{NSI} is to apply a Lorentz transformation to all of the final state partons: $\hat p_i = \Lambda p_i$ for $i = 1,\dots,m$. For this to work, we need
\begin{equation}
\label{eq:transformcondition}
(\hat p_\La + p_\Lb - \hat p_{m+1})^2
=
(p_\La + p_\Lb)^2
\;.
\end{equation}
In one way of proceeding, this condition determines $z$ in eq.~(\ref{hatpLa}) in terms of the three free components of $\hat p_{m+1}$. We will follow a slightly different alternative as follows. We need three splitting variables. Let one of them be $z$. Let the second be $y$. Let the third be the azimuthal angle $\phi$ of $\hat p_{m+1}$ around the beam axis.  Then eq.~(\ref{eq:transformcondition}) determines $\hat p_{m+1}$ as a function of $y$, $z$, and $\phi$.

To proceed with this program, we write $\hat p_{m+1}$ as
\begin{equation}
\label{eq:xaxbdef}
\hat p_{m+1} = 
x_\La n_\La
+ x_\Lb n_\Lb
+ k_\perp
\;.
\end{equation}
Here the direction of $k_\perp$ defines the azimuthal angle $\phi$ and the magnitude of $k_\perp$ is given by
\begin{equation}
\label{eq:kperpsq}
-k_\perp^2 = 
\left[
x_\La x_\Lb
- \hat\nu_{m+1}
\right]2 n_\La\cdot n_\Lb
\;.
\end{equation}
After a little bit of algebra, we find
\begin{equation}
\begin{split}
\label{eq:xaxbResult}
x_\La ={}&
\bigg[
\frac{1}{z}
-1 
- y \left(1+z\nu_\Lb\right)
- \left(1-z\right)\nu_\La \nu_\Lb  
- z\hat\nu_{m+1}\nu_\Lb 
\bigg]
\times\left[
1 - z^2
\hat\nu_\La\nu_\Lb
\right]^{-1}
\;,
\\
x_b ={}&
z
\bigg[y  + \hat\nu_{m+1} - \nu_\La
+(1+y)\,z\hat\nu_\La
+ z\hat\nu_\La\nu_\Lb
\left(\nu_\La 
- z 
\hat\nu_\La
\right)
\bigg]
\times\left[
1 - z^2
\hat\nu_\La\nu_\Lb
\right]^{-1}
\;.
\end{split}
\end{equation}
This decomposition of $\hat p_{m+1}$ is not exactly simple, but it is straightforward.

In the case of gluon emission, $\hat f_{m+1} =\Lg$, the matrix element is singular in the limit $(1-z) \to 0$. It is important in this case that there is a lower limit on $(1-z)$. When $\hat f_{m+1} = \Lg$, we have $\hat \nu_{m+1} = 0$ and $\hat \nu_a = \nu_\La$. Then the condition $x_\La > 0$ gives
\begin{equation}
(1-z) > y\, \frac{z (1 + z \nu_\Lb)}{1 - z \nu_\La \nu_\Lb}
\;,
\end{equation}
The exact condition is a little complicated, but with the approximations $y \ll 1$, $\nu_\La \ll 1$, and $\nu_\Lb \ll 1$ it is simple:
\begin{equation}
(1-z) \gtrsim y
\;.
\end{equation}
Notice that we do not assume any relation between $y$, $\nu_\La$, and $\nu_\Lb$, only that all three are small compared to 1.

We need the integration measure $d\zeta_\Lp$, which is given in ref.~\cite{NSI} eq.~(4.71):
\begin{equation}
d\zeta_\Lp = (2\pi)^{-3} d^4 \hat p_{m+1}\,
\delta(\hat p_{m+1}^2 - m^2(\hat f_{m+1}))\
\frac{\alpha+\beta/\eta_\La^2}{\hat\alpha + \hat\beta/\hat\eta_\La^2}
\;,
\end{equation}
where $\alpha,\beta,\hat\alpha$ and $\hat \beta$ are given in ref.~\cite{NSI}. 
When we introduce the variables $y$, $z$, $\phi$ and compute the Jacobian to $d^4 \hat p_{m+1}\,\delta(\hat p_{m+1}^2 - m^2(\hat f_{m+1}))$, we find
\begin{equation}
\begin{split}
\label{eq:dzetap}
d\zeta_\Lp ={}& 
\frac{2 n_\La \cdot n_\Lb}{4 (2\pi)^2}\,
\frac{1 - \nu_\La\nu_\Lb}
{1 - z^2 
\hat\nu_\La\,\nu_\Lb}\
dy\, \frac{dz}{z}\, \frac{d\phi}{2\pi}
\;.
\end{split}
\end{equation}
Notice that the mass dependent factor here is simply 1 in the limit $\nu_\La \ll 1$, $\hat \nu_\La \ll 1$, and $\nu_\Lb \ll 1$.

\subsection{Identifying the splitting function}
\label{sec:findinglambdaV}

With this information, we are prepared to identify splitting function $g_{a\hat a}$ in eq.~(\ref{eq:Veigenvalue}). From eq.~(\ref{eq:lambdaV1}), we have
\begin{equation}
\begin{split}
\label{eq:lambdaV2}
\lambda^{\cal V}_{\La\La}&(\{p,f\}_{m},t)
=
\\&
\frac{2 n_\La \cdot n_\Lb}{4 (2\pi)^2}
\sum_{\hat a}
\int\!dy \int\!\frac{dz}{z}\int\!\frac{d\phi}{2\pi}\
\frac{1 - \nu_\La\nu_\Lb}
{1 - z^2 \hat\nu_\La\,\nu_\Lb}\
\Theta(\zeta_{\rm p} \in \varGamma_{\La}(\{p\}_{m},\zeta_{\rm f}))
\\&\times
\delta\!\left(\log y  - \log\!\left(\frac{\eta_\La \mu_\LA^2}{2n_\La\cdot n_\Lb}
e^{-t}\right)
\right)
\\&\times
\frac
{n_\Lc(a)}
{n_\Lc(\hat a)}\,z\,
\frac{
\tilde f_{\hat a/A}(\eta_{\La}/z,\mu_\LA^2 e^{-t})}
{\tilde f_{a/A}(\eta_{\La},\mu_\LA^2 e^{-t})}
\,
C(\hat a, a)\,
\overline w_{\La\La}(\{\hat f,\hat p\}_{m+1})
\;.
\end{split}
\end{equation}
Here $\zeta_\Lp$ stands for the splitting variables and $\zeta_\Lp \in \Gamma_\La$ means that the variables are within their kinematic bounds. The bounds are determined by $0<x_\La$, $0 < x_\Lb$, and $\nu_{m+1} < x_\La x_\Lb$. The delta function that defines the shower time $t$ serves to eliminate the integration over $y$. Also, as we will see, $\overline w_{\La\La}$ does not depend on $\phi$ so we can immediately perform the integration over $\phi$. This gives
\begin{equation}
\begin{split}
\label{eq:lambdaV3}
\lambda^{\cal V}_{\La\La}(\{p,f\}_{m},t)
={}&
\frac{2 n_\La \cdot n_\Lb}{4 (2\pi)^2}
\sum_{\hat a}
\int\!dz\
\frac{1 - \nu_\La\nu_\Lb}
{1 - z^2 \hat\nu_\La\,\nu_\Lb}\
\theta(\zeta_{\rm p} \in \varGamma_{\La}(\{p\}_{m},\zeta_{\rm f}))
\\&\times
\frac
{n_\Lc(a)}
{n_\Lc(\hat a)}\,
\frac{
\tilde f_{\hat a/A}(\eta_{\La}/z,\mu_\LA^2 e^{-t})}
{\tilde f_{a/A}(\eta_{\La},\mu_\LA^2 e^{-t})}
\,
C(\hat a, a)\,y\,
\overline w_{\La\La}(\{\hat f,\hat p\}_{m+1})
\;,
\end{split}
\end{equation}
where we understand that
\begin{equation}
y = \frac{\eta_\La \mu_\LA^2}{2n_\La\cdot n_\Lb}\,
e^{-t}
\;.
\end{equation}
This enables us to identify the function $g_{a\hat a}$ in eq.~(\ref{eq:Veigenvalue}),
\begin{equation}
\begin{split}
\label{eq:lambdaV4}
\frac{\as}{2\pi}\,
\frac{1}{z}\,
g_{a\hat a}\!\left(z,{\mu^2}/{z},\{p,f\}_{m}\right)
={}&
\frac{2 n_\La \cdot n_\Lb}{4 (2\pi)^2}
\frac{1 - \nu_\La\nu_\Lb}
{1 - z^2 \hat\nu_\La\,\nu_\Lb}\
\Theta(\zeta_{\rm p} \in \varGamma_{\La}(\{p\}_{m},\zeta_{\rm f}))
\\&\times
\frac
{n_\Lc(a)}
{n_\Lc(\hat a)}\,
\,
C(\hat a, a)\,y\,
\overline w_{\La\La}(\{\hat f,\hat p\}_{m+1})
\;.
\end{split}
\end{equation}
Here $\mu^2$ us the virtuality in the splitting, $\mu^2 = y\, 2 n_\La\cdot n_\Lb$.

\subsection{The splitting functions $\overline w_{\La\La}$}
\label{sec:functionsbarw}

Let us begin with $(\hat a, a, f_{m+1}) = (q,q,\Lg)$. The spin averaged splitting function is given in eq.~(2.26) of ref.~\cite{NSII},
\begin{equation}
\begin{split}
\label{eq:waaquark}
\overline w_{\La\La} ={}& \frac{4\pi \alpha_\Ls}{2(n_\La\! \cdot\!  n_\Lb)^2}\,
\frac{1}{(2\,\hat p_\La\! \cdot\!  \hat p_\mpone)^2}\
D_{\mu\nu}(\hat p_\mpone,\hat Q)
\\&\times
\frac{1}{4}{\rm Tr}\left[
[\s{\hat p}_\La + m(a)]\gamma^\mu [\s{P}_\La + m(a)]\s{n}_\Lb
[\s{p}_\La + m(a)]\s{n}_\Lb[\s{P}_\La+ m(a)]\gamma^\nu
\right]
\;.
\end{split}
\end{equation}
Here $P_\La = \hat p_\La - \hat p_\mpone$ and $D_{\mu\nu}$ is the polarization sum for the emitted gluon, using timelike axial gauge with gauge fixing vector $\hat Q = \hat p_\La + p_\Lb$:
\begin{equation}
\label{eq:Dmunu}
D_{\mu\nu}(\hat p_\mpone,\hat Q) =
- g_{\mu\nu} 
+ \frac{\hat p_\mpone^\mu \hat Q^\nu + \hat Q^\mu \hat p_\mpone^\nu}
{\hat p_\mpone\cdot \hat Q}
- \frac{\hat Q^2 \hat p_\mpone^\mu \hat p_\mpone^\nu}{(\hat p_\mpone\cdot \hat Q)^2}
\;.
\end{equation}
The genesis of this splitting function is described in ref.~\cite{NSI}; evidently it is quite directly derived from the Feynman rules.

The function $\overline w_{\La \La}$ is a rather complicated function of $y$, $z$, $\nu_\La$  and $\nu_\Lb$, with $\hat\nu_\La = \nu_\La$ and $\hat\nu_{m+1} = 0$. However it is simple in the collinear limit $\lambda_\Lc \to 0$ with $y \propto \lambda_\Lc$, $\nu_\La \propto \lambda_\Lc$, $\nu_\Lb \propto \lambda_\Lc$. It is also simple in the soft limit, $\lambda_\Ls \to 0$ with $y \propto \lambda_\Ls$, $(1-z) \propto \lambda_\Ls$, $\nu_\La \propto \lambda_\Ls$, $\nu_\Lb \propto \lambda_\Ls$. A straightforward calculation gives the following form, which contains the leading behavior in both limits:
\begin{equation}
\begin{split}
\label{eq:waaqqg}
y z\, \overline w_{\La\La} \sim {}& 
\frac{8\pi\as}{2\,n_\La\! \cdot\!  n_\Lb}\
\left(
\frac{2}{1-z}
-(1+z)
-2 z \frac{\nu_\La}{y}
-\frac{2 y }{(1-z)^2}
\right)
\;.
\end{split}
\end{equation}
In the collinear limit, the first, second, and third terms in the parentheses are important. In the soft limit, the first and fourth terms in the parentheses are important. 

In eq.~(\ref{eq:lambdaV4}), there is also a theta function that gives the limits of integration over $y$ and $z$. This is, in the soft or collinear limits
\begin{equation}
\Theta(\zeta_{\rm p} \in \varGamma_{\La}(\{p\}_{m},\zeta_{\rm f}))
\sim \Theta((1-z) \nu_\La < y < (1-z))
\;.
\end{equation}
The restriction $(1-z) \nu_\La < y$, which applies in the collinear limit, comes from $x_\Lb > 0$. The restriction $y < (1-z)$, which applies in the soft limit, comes from $x_\La > 0$. 

Let us turn to $(\hat a, a, f_{m+1}) = (\Lg,\Lg,\Lg)$. The spin averaged splitting function is obtained by combining in Eqs.~(2.40), (2.45), (2.54), and (2.55) of ref.~\cite{NSII},
\begin{equation}
\begin{split}
\overline w_{\La\La} ={}&
\frac{{\pi\as}}{2 (\hat p_{\La}\!\cdot\! \hat p_{m+1})^2}\,
\\&\quad\times
v^{\alpha \beta \gamma}(\hat p_{m+1}, -\hat p_\La, \hat p_\La - \hat p_{m+1})\,
D_{\gamma\nu}(\hat p_\La - \hat p_{m+1}, n_\Lb)
\\&\quad\times
v^{\alpha' \beta' \gamma'}
(\hat p_{m+1}, -\hat p_\La, \hat p_\La - \hat p_{m+1})\,
D_{\gamma'\nu'}(\hat p_\La - \hat p_{m+1}, n_\Lb)
\\&\quad\times
D_{\alpha\alpha'}(\hat p_{m+1}, \hat  Q)\,
D_{\beta\beta'}(\hat p_{\La}, \hat Q)\,
D_{\nu\nu'}(p_{\La}, \hat Q)
\;.
\end{split}
\end{equation}
Here the three gluon vertex is
\begin{equation}
\label{eq:vgg}
v^{\alpha \beta \gamma}(p_a, p_b, p_c)
= g^{\alpha\beta} (p_a - p_b)^\gamma
+ g^{\beta\gamma} (p_b - p_c)^\alpha
+ g^{\gamma\alpha} (p_c - p_a)^\beta
\end{equation}
and $D_{\mu\nu}(p,\hat Q)$ is given in eq.~(\ref{eq:Dmunu}). The numerator function $D_{\gamma\nu}(\hat p_\La - \hat p_{m+1};n_l)$ projects onto the physical polarization states for the off-shell gluon. It is given by eq.~(\ref{eq:Dmunu}), where now the gauge fixing vector is $n_\Lb$, the lightlike vector in the direction of hadron B, as in the quark splitting function in eq.~(\ref{eq:waaquark}). The genesis of this splitting function is described in ref.~\cite{NSI}; evidently it is quite directly derived from the Feynman rules.

The function $\overline w_{\La \La}$ is a rather complicated function of $y$, $z$, and $\nu_\Lb$. However it is simple in the collinear limit $\lambda_\Lc \to 0$ with $y \propto \lambda_\Lc$, $\nu_\Lb \propto \lambda_\Lc$. It is also simple in the soft limit, $\lambda_\Ls \to 0$ with $y \propto \lambda_\Ls$, $(1-z) \propto \lambda_\Ls$, $\nu_\Lb \propto \lambda_\Ls$. A straightforward calculation gives the following form, which contains the leading behavior in both limits:
\begin{equation}
\begin{split}
\label{eq:waaggg}
y z\, \overline w_{\La\La} \sim {}& 
\frac{16\pi\as}{2\,n_\La\! \cdot\!  n_\Lb}\
\left(
\frac{1}{1-z}
+\frac{1}{z}
-2
+z(1-z)
-\frac{y}{(1-z)^2}
\right)
\;.
\end{split}
\end{equation}
In the collinear limit, the first four terms in the parentheses are important. In the soft limit, the first and fifth terms in the parentheses are important. 

In eq.~(\ref{eq:lambdaV4}), there is also a theta function that gives the limits of integration over $y$ and $z$. This is, in the soft or collinear limits
\begin{equation}
\Theta(\zeta_{\rm p} \in \varGamma_{\La}(\{p\}_{m},\zeta_{\rm f}))
= \Theta(y < (1-z))
\;.
\end{equation}
The restriction $y < (1-z)$, which applies in the soft limit, comes from $x_\La > 0$.

Let us turn next to $(\hat a, a, f_{m+1}) = (\Lg,q,\bar q)$. The spin averaged splitting function, derived from the definitions in ref.~\cite{NSI}, has a form similar that in eq.~(\ref{eq:waaquark}). In terms of dot products, it is given in eq.~(A.3) of ref.~\cite{NSII},
\begin{equation}
\begin{split}
\overline w_{\La\La} ={}&
\frac{{4\pi\as}}{\hat p_\La\!\cdot\! \hat p_{m+1}}
\left[
\frac{\hat p_\La\!\cdot\! n_\Lb}{p_\La\!\cdot \!n_\Lb}
-\frac{(\hat p_\La - \hat p_{m+1})\!\cdot\! n_\Lb}{p_\La\!\cdot \!n_\Lb}\,
\frac{\hat p_{m+1}^\mu\, D_{\mu\nu}(\hat p_\La,\hat Q)\,\hat p_{m+1}^\nu}
{\hat p_\La\!\cdot \!\hat p_{m+1}}
\right]
\;.
\end{split}
\end{equation}
The function $\overline w_{\La \La}$ is a fairly simple function of $y$, $z$, $\nu_\La$, and $\nu_\Lb$, with $\hat \nu_\La = 0$ and $\hat \nu_{m+1} = \nu_\La$. It is even simpler
 in the collinear limit $\lambda_\Lc \to 0$ with $y \propto \lambda_\Lc$, $\nu_\La \propto \lambda_\Lc$, $\nu_\Lb \propto \lambda_\Lc$. It is {\em not} more singular in the soft limit, $\lambda_\Ls \to 0$ with $y \propto \lambda_\Ls$, $(1-z) \propto \lambda_\Ls$, $\nu_\La \propto \lambda_\Ls$, $\nu_\Lb \propto \lambda_\Ls$. A straightforward calculation gives the following form, which contains the leading behavior in both limits:
\begin{equation}
\begin{split}
\label{eq:waaggg}
y z\, \overline w_{\La\La} \sim {}& 
\frac{8\pi\as}{2\,n_\La\! \cdot\!  n_\Lb}\
\left(
1 - 2 z (1-z) 
+ 2 z\,\frac{\nu_\La}{y}
\right)
\;.
\end{split}
\end{equation}

We also need the theta function in eq.~(\ref{eq:lambdaV4}) that gives the limits of integration over $y$ and $z$. In the soft limit, there is a lower bound on $(1-z)$ that comes from the requirement $x_\La > 0$: $(1-z) > y$. However, this bound is not relevant because there is no soft singularity in the case $(\hat a, a, f_{m+1}) = (\Lg,q,\bar q)$. There is, however, a restriction that arises from eq.~(\ref{eq:xaxbdef}). In order to have $|k_\perp^2|>0$, we need $x_\La x_\Lb > \nu_\La$. In the collinear limit, this gives $\nu_\La < y(1-z)$. Thus 
\begin{equation}
\Theta(\zeta_{\rm p} \in \varGamma_{\La}(\{p\}_{m},\zeta_{\rm f}))
\sim \Theta(\nu_\La < y(1-z))
\;.
\end{equation}

We turn finally to $(\hat a, a, f_{m+1}) = (q,\Lg,q)$. The spin averaged splitting function, derived from the definitions in ref.~\cite{NSI}, has a form similar that in eq.~(\ref{eq:waaquark}). In terms of dot products, it is given in eq.~(A.2) of ref.~\cite{NSII},
\begin{equation}
\begin{split}
\overline w_{\La\La} ={}&
\frac{{8\pi\as}}{(\hat p_\La - \hat p_{m+1})^2}
\left[
-1
+\left(\frac{\hat p_\La\!\cdot \!n_\Lb}{(\hat p_\La - \hat p_{m+1})\!\cdot\! n_\Lb}\right)^{\!2}
\frac{2\,\hat p_{m+1}^\mu\, D_{\mu\nu}(p_\La,\hat Q)\,\hat p_{m+1}^\nu}
{(\hat p_\La - \hat p_{m+1})^2}
\right]
\;.
\end{split}
\end{equation}
The function $\overline w_{\La \La}$ is a fairly simple function of $y$, $z$, $\hat\nu_\La$, and $\nu_\Lb$, with $\nu_\La = 0$ and $\hat \nu_{m+1} = \hat\nu_\La$. It is even simpler in the collinear limit $\lambda_\Lc \to 0$ with $y \propto \lambda_\Lc$, $\hat\nu_\La \propto \lambda_\Lc$, $\nu_\Lb \propto \lambda_\Lc$. It is {\em not} more singular in the soft limit, $\lambda_\Ls \to 0$ with $y \propto \lambda_\Ls$, $(1-z) \propto \lambda_\Ls$, $\hat\nu_\La \propto \lambda_\Ls$, $\nu_\Lb \propto \lambda_\Ls$. A straightforward calculation gives the following form, which contains the leading behavior in both limits:
\begin{equation}
\begin{split}
\label{eq:waaqgq}
y z\, \overline w_{\La\La} \sim {}& 
\frac{8\pi\as}{2\,n_\La\! \cdot\!  n_\Lb}\
\left(
\frac{1+(1-z)^2}{z}
- 2 z\,\frac{\hat\nu_\La}{y}
\right)
\;.
\end{split}
\end{equation}

We again need the theta function in eq.~(\ref{eq:lambdaV4}) that gives the limits of integration over $y$ and $z$. In the soft limit, there is the bound $(1-z) > y$. However, this bound is not relevant because there is no soft singularity in the case $(\hat a, a, f_{m+1}) = (q,\Lg, q)$. There is a restriction that arises from eq.~(\ref{eq:xaxbdef}): $x_\La x_\Lb > \hat\nu_a$. In the collinear limit, this gives $z^2 \hat\nu_\La < y(1-z)$. Thus 
\begin{equation}
\Theta(\zeta_{\rm p} \in \varGamma_{\La}(\{p\}_{m},\zeta_{\rm f}))
\sim \Theta(z^2 \hat\nu_\La < y(1-z))
\;.
\end{equation}

\subsection{The approximate splitting functions}

We can now use eq.~(\ref{eq:lambdaV4}) to identify the functions $g_{a\hat a}$ in eq.~(\ref{eq:Veigenvalue}). We are interested in the lea{}ding $y\to 0$ behavior, accounting for both the collinear limit and the soft limit. In these limits, we have
\begin{equation}
\begin{split}
\label{eq:lambdaV4limit}
\frac{\as}{2\pi}\,
g_{a\hat a}\!\left(z,{\mu^2}/{z},\{p,f\}_{m}\right)
={}&
\frac
{n_\Lc(a)}
{n_\Lc(\hat a)}\,
\,
C(\hat a, a)\,
\Theta(\zeta_{\rm p} \in \varGamma_{\La}(\{p\}_{m},\zeta_{\rm f}))
\\&\times
\frac{2 n_\La \cdot n_\Lb}{4 (2\pi)^2}\,yz\,
\overline w_{\La\La}(\{\hat f,\hat p\}_{m+1})
\;.
\end{split}
\end{equation}
Here we use the limiting forms for $\overline w_{\La\La}$ and for the theta function that we worked out in the previous section. Using our results, $g_{a\hat a}$ has the form
\begin{equation}
\begin{split}
\label{eq:gtoGbis}
g_{a\hat a}\!\left(z,{\mu^2}/{z},\{p,f\}_{m}\right)
\sim{}& 
\bigg[
G_{a\hat a}\!\left(z,{\mu^2}/{z}\right)
- \delta_{a\hat a}\,\frac{2C_a\,y}{(1-z)^2}
\bigg]\Theta(y <(1-z))
\;,
\end{split}
\end{equation}
where
\begin{equation}
\begin{split}
G_{qq} ={}& C_\LF\left[\frac{2}{1-z}
-(1+z)
-2 z\,\frac{\nu_\La}{y}
\right]\Theta((1-z) \nu_\La < y)
\;,\\
G_{gg} ={}& 2C_\LA
\left[\frac{1}{1-z}
+\frac{1}{z}
-2
+z(1-z)
\right]
\;,\\
G_{qg} ={}& T_\LR\left[1 - 2 z (1-z) 
+ 2 z\,\frac{\nu_\La}{y}\right]
\Theta(\nu_\La < y(1-z))
\;,\\
G_{gq} ={}& C_\LF\left[\frac{1+(1-z)^2}{z}
- 2 z\,\frac{\hat\nu_\La}{y}\right]
\Theta(z^2 \hat\nu_\La < y(1-z))
\;.
\end{split}
\end{equation}
These are the results used in eq.~(\ref{eq:gsoftregion}) for the soft limit and eq.~(\ref{eq:kernels}) for the collinear limit.

\end{document}